\documentclass[twocolumn]{aastex63}
\usepackage{verbatim}

\received{XXXX, 2020}
\revised{XXXX, 2020}
\accepted{January 31, 2022}
\submitjournal{ApJ}

\shorttitle{Finding IMBHs in the optical and mid-IR}
\shortauthors{Richardson et al.}

\begin{document}

\title{Optical and JWST Mid-IR Emission Line Diagnostics for Simultaneous IMBH and Stellar Excitation in $z\sim0$ Dwarf Galaxies}

\correspondingauthor{Chris Richardson}
\email{crichardson17@elon.edu}

\author[0000-0002-3703-0719]{Chris T. Richardson}
\affiliation{Elon University \\
100 Campus Drive \\
Elon, NC 27278, USA}
\affiliation{COSmS Institute \\
University of North Carolina \\
Chapel Hill, NC 27599, USA}


\author{Connor Simpson}
\affiliation{Elon University \\
100 Campus Drive \\
Elon, NC 27278, USA}

\author[0000-0001-6162-3963]{Mugdha S. Polimera}
\affiliation{
University of North Carolina \\
141 Chapman Hall CB 3255 \\
Chapel Hill, NC 27599, USA}

\author{Sheila J. Kannappan}
\affiliation{
University of North Carolina \\
141 Chapman Hall CB 3255 \\
Chapel Hill, NC 27599, USA}

\author{Jillian M. Bellovary}
\affiliation{
Queensborough Community College\\
City University of New York\\
222-05 56th Ave\\
Bayside, NY 11364, USA\\}
\affiliation{
American Museum of Natural History\\
Central Park West at 79th Street\\
New York, NY 10024, USA}
\affiliation{
Graduate Center\\
City University of New York\\
New York, NY 10016, USA}

\author{Christopher Greene}
\affiliation{University of Cincinnati \\
290 CCM Blvd \\
Cincinnati, OH 45221, USA}

\author{Sam Jenkins}
\affiliation{Elon University \\
100 Campus Drive \\
Elon, NC 27278, USA}





\begin{abstract}

Current observational facilities have yet to conclusively detect $10^3 - 10^4 M_{\odot}$ intermediate mass black holes (IMBHs) that fill in the evolutionary gap between early universe seed black holes and $z \sim 0$ supermassive black holes. Dwarf galaxies present an opportunity to reveal active IMBHs amidst persistent star formation. We introduce photoionization simulations tailored to address key physical uncertainties: coincident vs. non-coincident mixing of IMBH and starlight excitation, open vs. closed surrounding gas cloud geometries, and different AGN SED shapes. We examine possible AGN emission line diagnostics in the optical and mid-IR, and find that the diagnostics are often degenerate with respect to the investigated physical uncertainties. In spite of these setbacks, and in contrast to recent work, we are able to show that [O~III]/H$\beta$ typically remains bright for dwarf AGN powered by IMBHs down to $10^3 M_{\odot}$. Dwarf AGN are predicted to have inconsistent star-forming and Seyfert/LINER classifications using the most common optical diagnostics. In the mid-IR, [O~IV] 25.9$\mu$m and [Ar~II] 6.98$\mu$m are less sensitive to physical uncertainties than are optical diagnostics. Based on these emission lines, we provide several mid-IR emission line diagnostic diagrams with demarcations for separating starbursts and AGN with varying levels of activity. The diagrams are valid over a wide range of ionization parameters and metallicities out to $z\sim0.1$, so will prove useful for future JWST observations of local dwarf AGN in the search for IMBHs. We make our photoionization simulation suite freely available.

\end{abstract}



\section{Introduction} \label{sec:intro}

The occupation fraction of supermassive black holes (SMBHs) in massive galaxies is near unity (\citealt{Magorrian1998}). While LIGO has detected stellar mass black holes ($\leq 10^2 M_{\odot}$) resulting from compact object mergers, intermediate black holes (IMBHs) remain elusive in the $10^2 M_{\odot}$ - $10^5 M_{\odot}$ range (\citealt{Greene2020}). Detecting black holes at the low end of this range would provide a crucial link between early universe black hole seeds and local SMBHs.

\begin{figure*}[ht!]
\includegraphics[width=2.15\columnwidth]{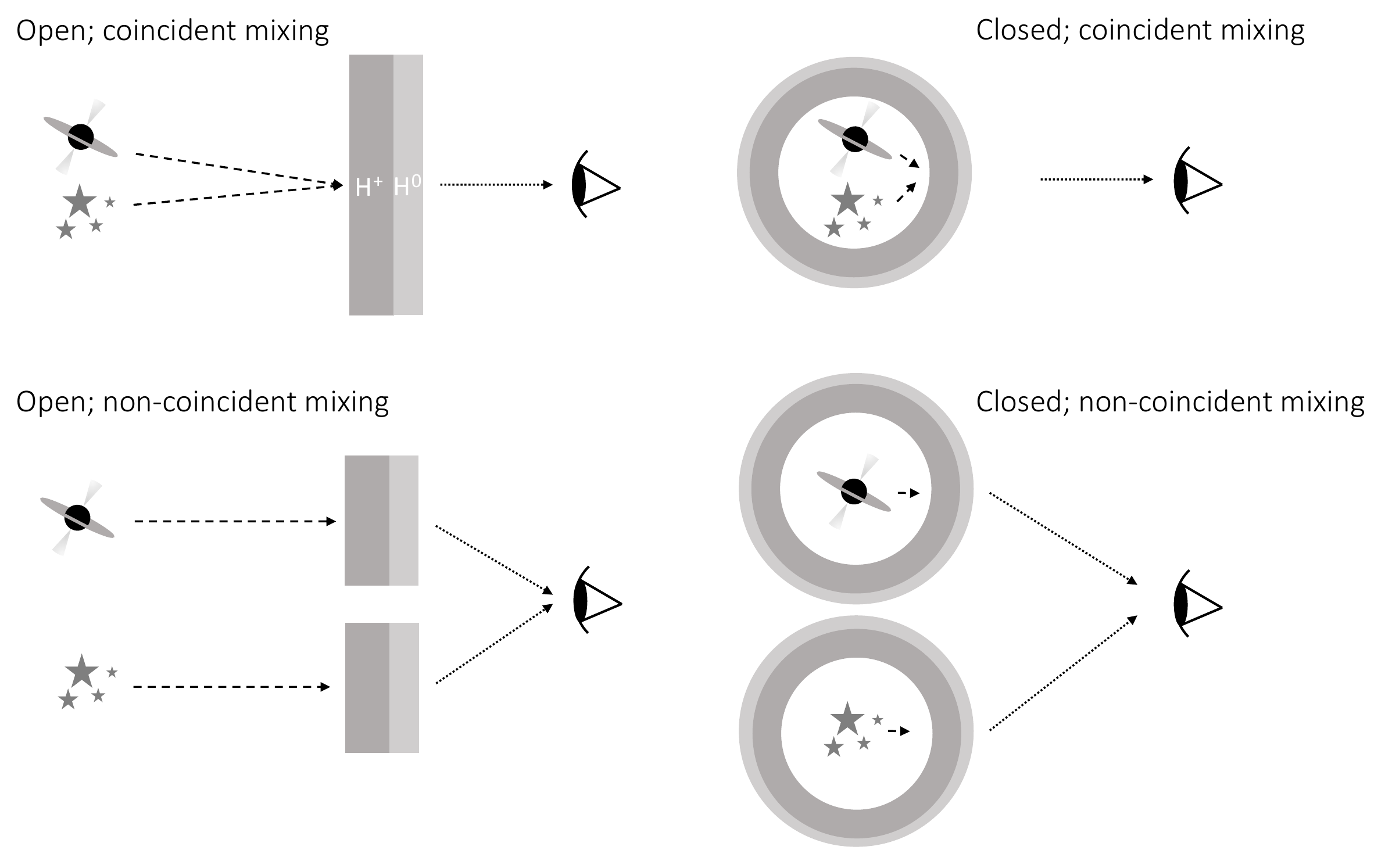}
\caption{Illustration depicting two different gas cloud geometries (open and closed) and two different mixing methodologies (coincident and non-coincident). Dwarf AGN could presumably fall into any of these four categories and therefore photoionization models must account for this systematic uncertainty. Note that the rays of light (arrows) only show one possible path to the observer and do not represent all possible components of the observed spectrum.\\
\label{fig:geometry}}
\end{figure*}

Dwarf galaxies show promise in the search for IMBHs based on the $M_*-M_{\mathrm{BH}}$ relation (\citealt{Reines+Volonteri2015}). However, the properties of dwarf AGN hosts are different from those of massive galaxy AGN hosts, which can complicate detecting AGN signatures. For example, most dwarfs are gas-rich (\citealt{Kannappan2004}), strongly star-forming (\citealt{Geha2012}), and metal-poor (\citealt{Tremonti2004}). Supernova feedback preferentially expels metal-rich gas (\citealt{MacLow1999}), while accretion of low metallicity, intergalactic medium gas drives star formation (\citealt{Dekel2006}).

Unlike in the unified AGN model, IMBHs in dwarfs often wander within 1 kpc of the center due to the dwarfs' weak gravitational potential (\citealt{Reines2020}, \citealt{Bellovary2021}). Therefore, it is unclear whether the AGN and stellar radiation fields strike the same gas clouds, or spatially separated gas clouds, as the IMBH relocates or settles down in a particular location. This uncertainty calls into question applying the centralized geometric model for massive AGN to dwarf AGN.

Another issue for dwarf AGN, unlike massive AGN, is that X-ray observations that could provide valuable constraints on the AGN spectral energy distribution (SED) remain rare for $M_{\mathrm{BH}} \approx 10^5 M_{\odot}$, and absent for $M_{\mathrm{BH}} \approx 10^3 M_{\odot}$ (\citealt{Desroches2009}). The few sources luminous enough to observe are likely outliers having fortuitous conditions to enable their detection (e.g. \citealt{Godet2012}). This situation introduces doubts about whether physical quantities controlling the shape of the SED scale down to the lowest black hole masses (\citealt{Arcodia2020}).

Photoionization models can provide the missing ingredient for IMBH detection by systematically accounting for uncertainties pertaining to the gaseous geometry and properties of low-mass AGN. The left panels of Figure \ref{fig:geometry} illustrate open, or plane-parallel geometries, where the covering factor is small, as typically assumed when modeling AGN (\citealt{Elvis2000}; \citealt{Feltre2016}) and star-forming regions like the Orion blister (\citealt{Ferland2001}). The right panels of Figure \ref{fig:geometry} illustrate closed, or spherical geometries, where the covering factor is close to unity, as typically assumed when modeling obscured AGN like ULIRGs (\citealt{Abel2009}) and star-forming regions like 30 Doradus (\citealt{Pellegrini2011}). 

Similarly, photoionization models have used two methods to take into account simultaneous AGN and stellar excitation, but rarely with justification. Figure \ref{fig:geometry}, top panels, illustrates one approach where an AGN and starlight strike the same cloud and thus the SEDs from each source are mixed \textit{a priori} (e.g. \citealt{Abel2009}, \citealt{Satyapal2018}), which we label as \textit{coincident mixing}. The bottom panels of Figure \ref{fig:geometry} show another approach where the excitation sources illuminate spatially separated clouds and thus the models are mixed \textit{a posteriori} (e.g. \citealt{Kewley2013}, \citealt{Melendez2014}, \citealt{Richardson2014}), which we label as \textit{non-coincident mixing}. Since dwarfs hosting IMBHs display a variety of morphologies (\citealt{Kimbrell2021}), it is unclear whether a particular gas geometry or mixing methodology would generically apply. Therefore, assessing all possibilities in Figure~\ref{fig:geometry} is paramount.

Optical spectroscopy enables the categorization of emission line galaxies as AGN, star-forming, or a mixture of the two using [O~III] $\lambda$5007/H$\beta$ against [N~II] $\lambda$6584/H$\alpha$ (i.e., the BPT diagram), [S~II] $\lambda$6720/H$\alpha$, and [O~I] $\lambda$6300/H$\alpha$ to form diagnostic diagrams (\citealt{Baldwin1981}, \citealt{V&O87}). Previous photoionization modeling including AGN suggests that optical lines might grow too faint for detection for black holes outside $10^6 - 10^9 M_{\odot}$, thus skewing $M_{\mathrm{BH}}$ distributions (\citealt{Cann2019}, \citealt{Bhat2020}); however, none of this modeling accounts for the multiple geometrical configurations indicated in Figure~\ref{fig:geometry}, for the presence of stellar excitation, or for the uncertainty in the shape of the IMBH SED. 

Indeed, optical observations may contradict the theoretical impossibility of detection of IMBH AGN. Broad-line selected dwarf AGN with $M_{\mathrm{BH}} \approx 10^{5} M_{\odot}$ can be sometimes optically classified as AGN using the BPT diagram (\citealt{Barth2004}), indicating high [O~III]/H$\beta$. \cite{Reines2020} used radio interferometry to identify dwarf AGN with optical star-forming galaxies classifications, and used the $M_*-M_{\mathrm{BH}}$ relation to deduce $M_{\mathrm{BH}} \sim 10^{4.1}-10^{5.8} M_{\odot}$, albeit this relation shows up to 1.0 dex scatter for $M_*<10^9 M_{\odot}$ (\citealt{Greene2020}). Optically classified star-forming galaxies might contain a treasure trove of additional hidden IMBHs that evade detection on account of the BPT diagram preferentially identifying high metallicity AGN (Polimera et al., submitted; hereafter P21). However, active black holes in the $10^3~M_{\odot}$ regime still remain undetected.

IR spectroscopy offers a better opportunity to reliably detect IMBHs as it possesses a wealth of high ionization lines insensitive to gas metallicity and dust extinction. The Spitzer era revealed the potential of the mid-IR to separate starbursts from AGN, leading to the development of several diagnostic diagrams involving [Ne~V] and [O~IV] emission lines and polycyclic aromatic hydrocarbon (PAH) features (\citealt{Dale2006}). The James Webb Space Telescope (JWST), spanning $0.6 - 28.3~\mu m$, has the potential to revolutionize the search for black holes at the low-mass end of the IMBH distribution. Indeed, recent work has shown the  spectral range of JWST can uncover AGN eluding detection from optical spectroscopy (\citealt{Satyapal2020}), although the uncertainties stemming the from IMBH SED and the configurations presented in Figure \ref{fig:geometry} remain unexplored.

In this paper, we fill in the gap in photoionization modeling to account for the uncertainty in gaseous geometry, mixed excitation, and AGN SED shape. We make emission line predictions that will be valuable for searching for $10^3 - 10^5 M_{\odot}$ black holes with optical spectroscopy and future JWST observations, while freely providing our simulation suite to the community\footnote{\url{https://facstaff.elon.edu/crichardson17/}}.

\section{Theoretical Methodology} \label{sec:method}

\subsection{Incident Radiation Field}

\begin{figure}[t!]
\includegraphics[width=1.0\columnwidth]{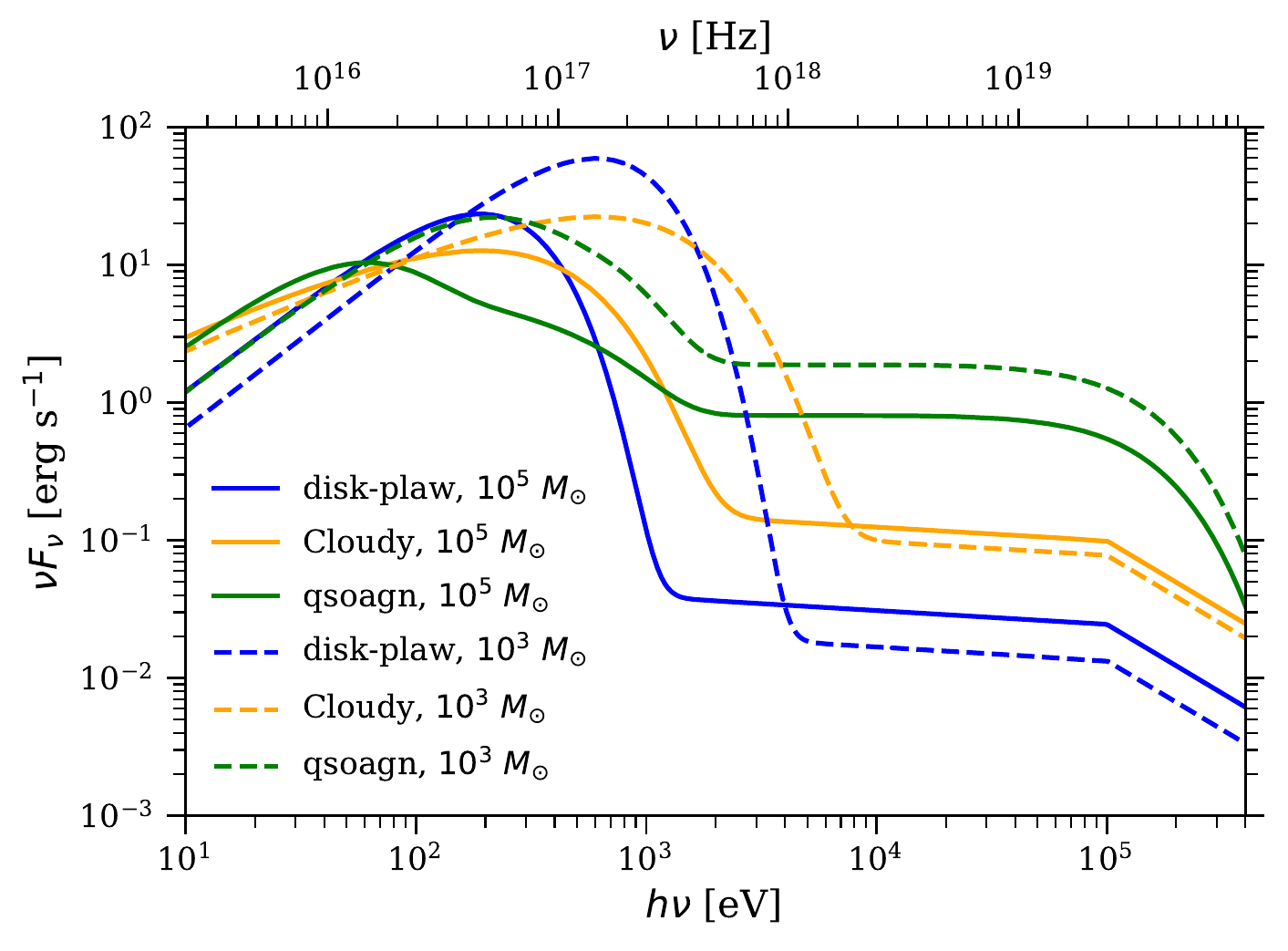}
\caption{AGN SEDs resulting from different models for $M_{\mathrm{BH}} = 10^{3} - 10^{5} ~M_{\odot}$. Profound differences are present for the peak of the thermal accretion disk and the hard X-ray contribution.
\label{fig:seds}}
\end{figure}

Figure \ref{fig:seds} displays the three different models for the AGN SED that we have explored, assuming $M_{\mathrm{BH}} = 10^3 - 10^5 M_{\odot}$: ``disk-plaw," ``Cloudy," and ``\texttt{qsosed}." The ``disk-plaw" SED combines the \texttt{diskbb} accretion disk model (\citealt{Mitsuda1984}) with a power law ($\Gamma = 2.1$), normalized to give $\alpha_{ox} = 1.41$ (\citealt{Grupe2010}) where

\begin{equation}
\frac{f_{\nu}(2~\mathrm{keV})}{f_{\nu}(2500 \AA)} = \left( \frac{\nu_{2~\mathrm{keV}}}{\nu_{2500 \AA}} \right)^{\alpha_{ox}}.
\end{equation}

\noindent The inner \texttt{diskbb} temperature is calculated using \cite{Peterson1997},

\begin{equation}
\small T_{\mathrm{in}} = 6.3 \times 10^5 \left( \frac{\dot{m}}{\dot{m}_{\mathrm{Edd}}}  \right)^{1/4}  \left( \frac{M_{\mathrm{BH}}}{10^8 ~ M_{\odot}} \right)^{-1/4} \left( \frac{R}{R_s} \right)^{-3/4} ~~ \mathrm{K}
\label{eqn:T_in}
\end{equation}

\noindent where we assume $\dot{m}/\dot{m}_{\mathrm{Edd}} = 0.1$ and $R = 3R_S$. The result is a piecewise, physically motivated SED for photoionization modeling (\citealt{Cann2019}, \citealt{Bhat2020}), but lacks physical self-consistency, such as the accretion disk radiation generating seed photons for the hard X-ray component. 

The ``Cloudy" SED is the \texttt{Cloudy} (\citealt{Ferland2017}) default AGN SED, which is an empirical model that assumes the observed SED is the same as the continuum seen by nebular clouds. The functional form of the SED is given by,

\begin{equation}
f_{\nu} = \nu^{\alpha_{\mathrm{uv}}} e^{\frac{-h\nu}{kT_{\mathrm{peak}}}} e^{\frac{-kT_{\mathrm{IR}}}{h\nu}} + a \nu^{\alpha_x}
\end{equation}

\noindent where $T_{\mathrm{peak}} = 0.77 ~ T_{\mathrm{in}}$ (\citealt{Mitsuda1984}), $\alpha_{\mathrm{UV}}$ is the low energy slope in the UV, $\alpha_{\mathrm{X}}$ is X-ray slope, and $a$ is a constant adjusted to satisfy $\alpha_{\mathrm{ox}}$. All of the spectral indices are taken from the median of the extinction corrected BLS1s sample in \cite{Grupe2010}.

The ``\texttt{qsosed}" uses the \texttt{agnsed} model (\citealt{Kubota2018}), while fixing most parameters to their ``typical" values and assuming $\dot{m}/\dot{m}_{\mathrm{Edd}} = 0.1$, $a_* = 0$, and $i=45^{\circ}$. This is a physically self-consistent SED for photoionization modeling (\citealt{Panda2019}, \citealt{Vasiliev2020}, \citealt{Sarkar2021}) appropriate for sub-Eddington accretion in IMBHs.

\begin{figure}[t!]
\includegraphics[width=1.0\columnwidth]{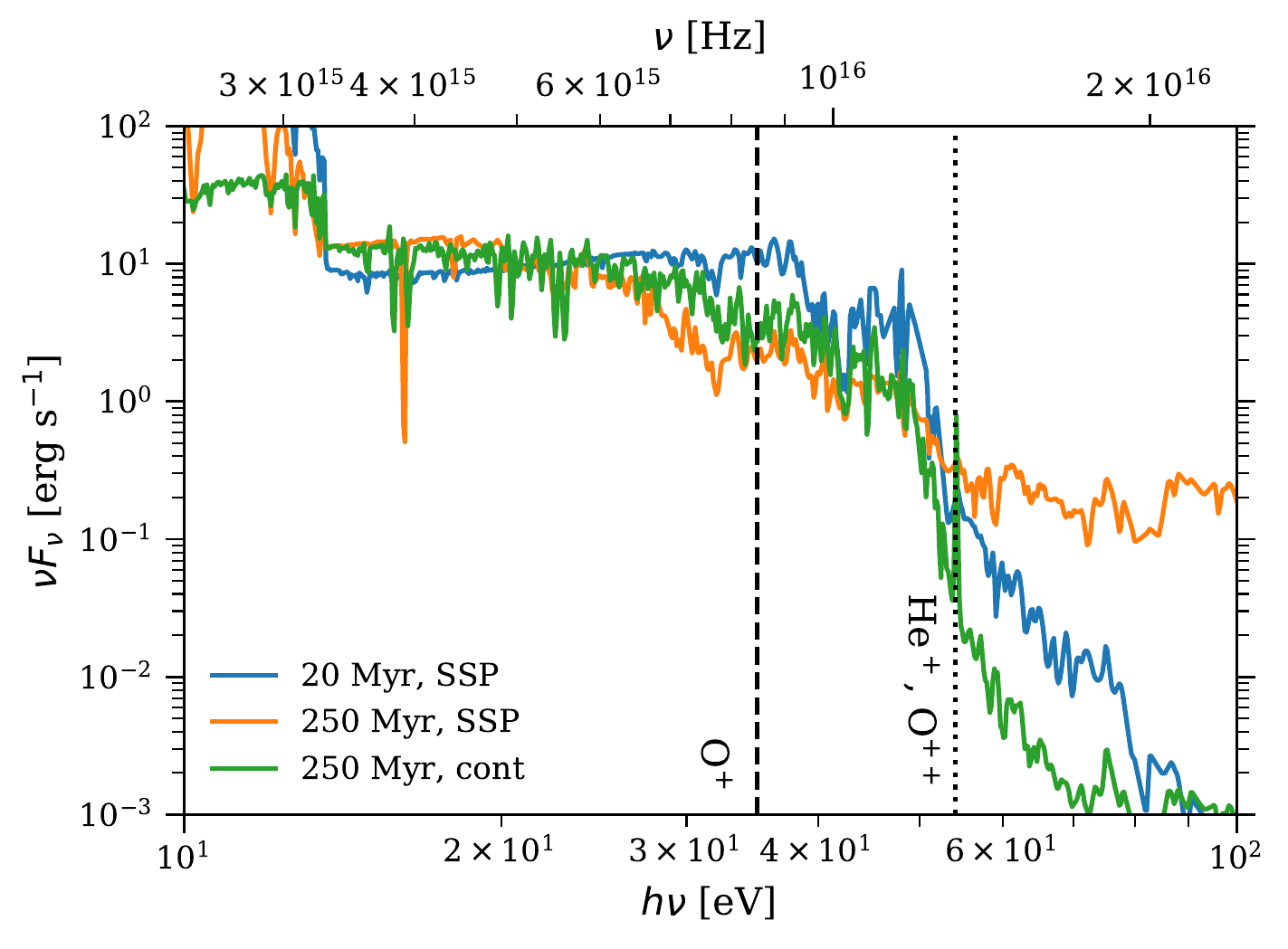}
\caption{Stellar SEDs resulting from different SFHs. The ionization potentials for relevant species are denoted by vertical lines.
\label{fig:stellar_seds}}
\end{figure}

\begin{table}
	\centering
	\caption{Reference abundances and depletion factors $\delta$ used for included chemical elements $X$. We set non-refractory element He, Ne, S, and Ar to  $\delta_X$ = 0. See Appendix A for details.}
	\label{tab:abundance_table}
	\begin{tabular}{lcr} 
		\hline
		X & log $X/$H & $\delta_X (F_* = 0.45)$ \\
		\hline
		He & -1.01 & 0.0 \\
		Li & -8.722 & -0.524 \\
		Be & -10.68 & -0.274 \\
		B & -9.193 & -0.546 \\
		C & -3.577 & -0.120 \\
		N & -4.21 & 0.000 \\
		O & -3.24 & -0.112 \\
		F & -7.56 & -0.147 \\
		Ne & -3.91 & 0.0 \\
		Na & -5.79 & -0.538 \\
		Mg & -4.44 & -0.659 \\
		Al & -5.57 & -1.602 \\
		Si & -4.50 & -0.625 \\
		P & -6.59 & 0.000 \\
		S & -4.88 & 0.0 \\
		Cl & -6.75 & -0.037 \\
		Ar & -5.60 & 0.0 \\
		K & -6.96 & -0.614 \\
		Ca & -5.68 & -2.356 \\
		Sc & -8.84 & -1.533 \\
		Ti & -7.07 & -1.928 \\
		V & -8.11 & -1.159 \\
		Cr & -6.38 & -1.379 \\
		Mn & -6.58 & -1.134 \\
		Fe & -4.48 & -1.510 \\
		Co & -7.07 & -1.343 \\
		Ni & -5.80 & -1.517 \\
		Cu & -7.82 & -0.757 \\
		Zn & -7.44 & -0.075 \\
		\hline
	\end{tabular}
\end{table}

Figure \ref{fig:seds} shows profound differences between the three SED models. In part, the differences are due to the self-consistent physics in the \texttt{qsosed} that is not featured in the other two models. The \texttt{qsosed} selects the inner accretion disk radius needed to power the X-ray emission and then truncates the disk at that point, rather than arbitrarily selecting $R_{in}=6GM/c^2$. This results in the \texttt{qsosed} SED for $M_{\mathrm{BH}} = 10^{3}~M_{\odot}$ having approximately the same peak energy as the other two SEDs for $M_{\mathrm{BH}} = 10^{5}~M_{\odot}$, and also a more appreciable hard X-ray component. Throughout the remainder of the paper, we only use the two SEDs that represent extremes: disk-plaw and \texttt{qsosed}.


To model the starburst continuum, we use the binary stellar population synthesis (SPS) code BPASS v2.0 (\citealt{Stanway2016}) to predict the spectrum emitted from stars subject to binary evolution. Including binaries is essential for a realistic treatment of the Wolf-Rayet (WR) phase, which can result from mergers and envelope removal (\citealt{D'Agostino2019}). As outlined in \cite{Richardson2019}, we adopt a Kroupa IMF with exponents 1.3 and 2.35 over the mass ranges $0.1 M_{\odot} < M < 0.5 M_{\odot}$ and $0.5 M_{\odot} < M < 300 M_{\odot}$ with one of three star formation histories (SFHs): an instantaneous burst SFH at age 20~Myr, an instantaneous burst SFH at age 250~Myr, or a continuous SFH at age 250~Myr. For a single stellar population (SSP), 20~Myr corresponds to the age where [O~III]/H$\beta$ reaches a maximum (\citealt{Xiao2018}), while 250 Myr corresponds to the age where the ionizing continuum flux for He$^{+}$ and O$^{++}$ ($>$54~eV) reaches a maximum. After continuous star formation for 250 Myr, the ionizing continuum ceases to evolve and this provides a ``general" use stellar SED for photoionization modeling. We include 11 different metallicities calibrated to Z$_{\odot}$~=~0.02 spanning 0.05 Z$_{\odot}$ -- 2.0 Z$_{\odot}$ for all SFHs. Figure \ref{fig:stellar_seds} displays stellar SEDs for each of the three SFHs at 0.4~Z$_{\odot}$.

To mix the AGN and stellar SEDs, we use both coincident and non-coincident mixing (Figure \ref{fig:geometry}) with AGN fractions ($f_{\mathrm{AGN}}$) = 0.0, 0.04, 0.08, 0.16, 0.32, 0.5, 0.64, and 1.0, where $f_{\mathrm{AGN}}$ represents the fraction of the total ionizing continuum attributed to the AGN SED. Additionally, the cosmic ray background value $\xi = 2.0 \times 10^{-16} ~ \mathrm{s}^{-1}$ (\citealt{Indriolo2007}) is added to satisfy the chemistry network.

\subsection{Gaseous Cloud \label{sec:gas-cloud}}

Following \cite{Richardson2019}, we select a hydrogen density of log $n_{\mathrm{H}}$ = 2.0 [cm$^{-3}$] at the illuminated face. After selecting $n_{\mathrm{H}}$, the ionization parameter is then given by,

\begin{equation}
    U = \frac{\phi_{\mathrm{H}}}{n_{\mathrm{H}} c}
	\label{eq:U}
\end{equation}

\noindent where $\phi_{\mathrm{H}}$ is the hydrogen ionizing flux. While most emission line galaxies indicate log $U$ = -3.5 -- -2.0, an even lower limit is needed to explain the lowest ionization dwarf galaxies and log $U$ $>$ -1.5 is needed to explain local blue compact dwarf galaxies (\citealt{Stasinska2015}). Accordingly, we run simulations with ionization parameters from log $U$ = -4.0 to log $U$ = -0.5 in increments of $\Delta$(log $U$) = 0.25. We follow \cite{Abel2008} employing a magnetic field and constant pressure equation of state until all simulations stop at $n_e/n_{\mathrm{H}} = 0.01$. We include a small amount of turbulence ($v=2$~km/s) to reduce line trapping. We consider both open and closed geometries as given in Figure \ref{fig:geometry}.

We use the methodology in \cite{Nicholls2017} for our abundances and scaling. The solar values for Galactic Concordance abundances are largely based off of \cite{Nieva2012}, \cite{Grevesse2015}, and \cite{Scott2015b,Scott2015a}, which we list in Table \ref{tab:abundance_table}. The scaling of abundances with metallicity includes a detailed prescription for accounting for specific elemental variations due to differences in nucleosynthesis. Unfortunately, the abundance of oxygen relative to solar has become synonymous with metallicity. While metallicity is strictly defined as the mass fraction of metals, and oxygen makes the greatest contribution to metallicity, the two are not equivalent. To avoid this ambiguity, we refer to the scaling parameter $\zeta_\mathrm{O}$ as the metallicity (see \citealt{Nicholls2017}) where the solar metallicity is 12+log(O/H) = 8.76, corresponding to $\zeta_\mathrm{O} = 1$.

We assume Orion grains and PAHs throughout the cloud as implemented in \cite{Baldwin1991} and \cite{Abel2008}, respectively. The dust abundance is typically assumed to not depend on metallicity, while the gas (hydrogen) to dust ratio (G/D) varies as G/D~$\propto Z^{-1}$ (\citealt{Dwek1998}). Here, we adopt a more sophisticated broken power law,

\begin{equation}
\mathrm{log~G/D} = \left\{
        \begin{array}{ll}
            2.21 - \mathrm{log} \frac{Z_{\mathrm{gas}}}{Z_{\odot}} & \quad Z_{\mathrm{gas}} \geq 0.25  Z_{\odot} \\\\
            0.96 - 3.10 ~ \mathrm{log} \frac{Z_{\mathrm{gas}}}{Z_{\odot}} & \quad Z_{\mathrm{gas}} < 0.25  Z_{\odot}
        \end{array}
    \right.
\end{equation}

\noindent which shows that low metallicity dwarfs deviate from a single power law relation (\citealt{Remy-Ruyer2014}). The Orion grain abundances and PAH abundances are scaled from their default abundances by the same factor to satisfy this relationship at a given $Z$.

\begin{figure*}[t!]
\includegraphics[width=2.15\columnwidth]{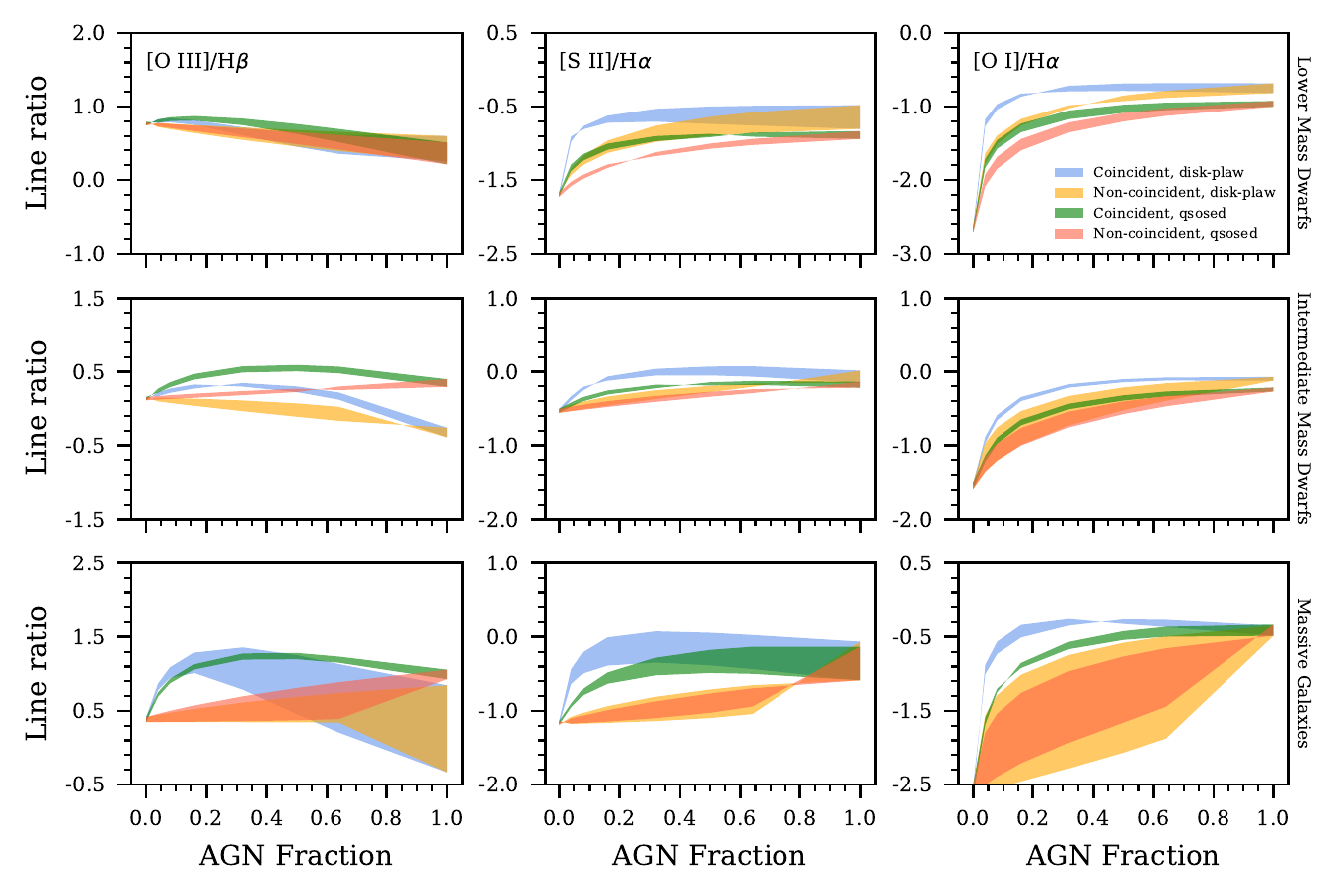}
\caption{Common optical emission line excitation diagnostics in different galaxy mass regimes (right label) assuming $M_{\mathrm{BH}} = 10^{3}~M_{\odot}$. The width of the line represents uncertainty due to geometry, while the colors refer to the mixing methodologies from Figure \ref{fig:geometry} and two SEDs from Figure \ref{fig:seds}. Each line has been thickened by 0.04 dex for clarity and the y-axis of each panel spans a 3.0 dex range to highlight the relative sensitivity of each emission line to the physical parameters explored.
\label{fig:AGN-frac-optical}}
\end{figure*}

\begin{figure*}[t!]
\includegraphics[width=2.15\columnwidth]{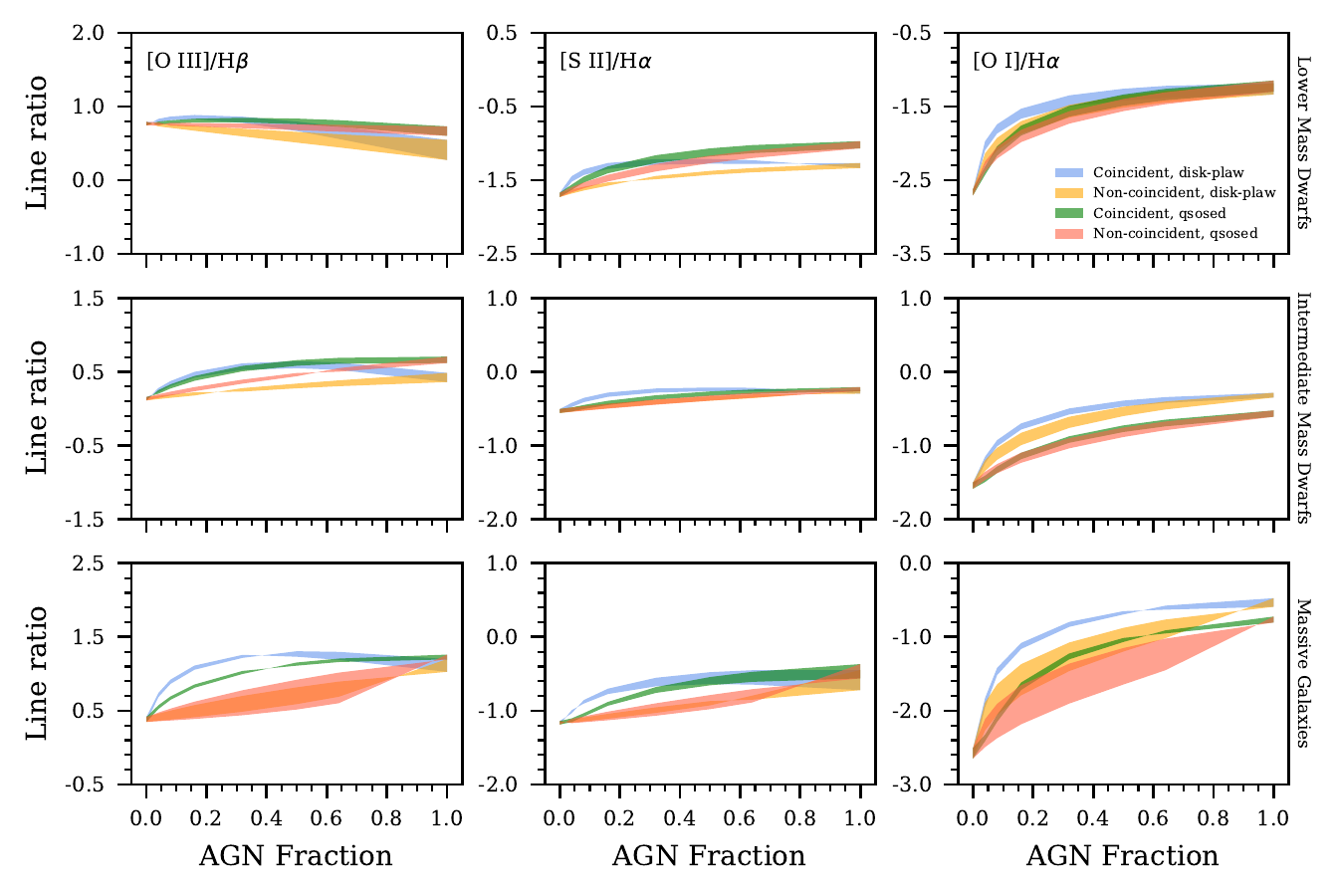}
\caption{Common optical emission line excitation diagnostics in the same manner as Figure \ref{fig:AGN-frac-optical} except assuming $M_{\mathrm{BH}} = 10^{5}~M_{\odot}$.
\label{fig:AGN-frac-optical_M-5_0}}
\end{figure*}

\begin{figure*}[t!]
\includegraphics[width=2.15\columnwidth]{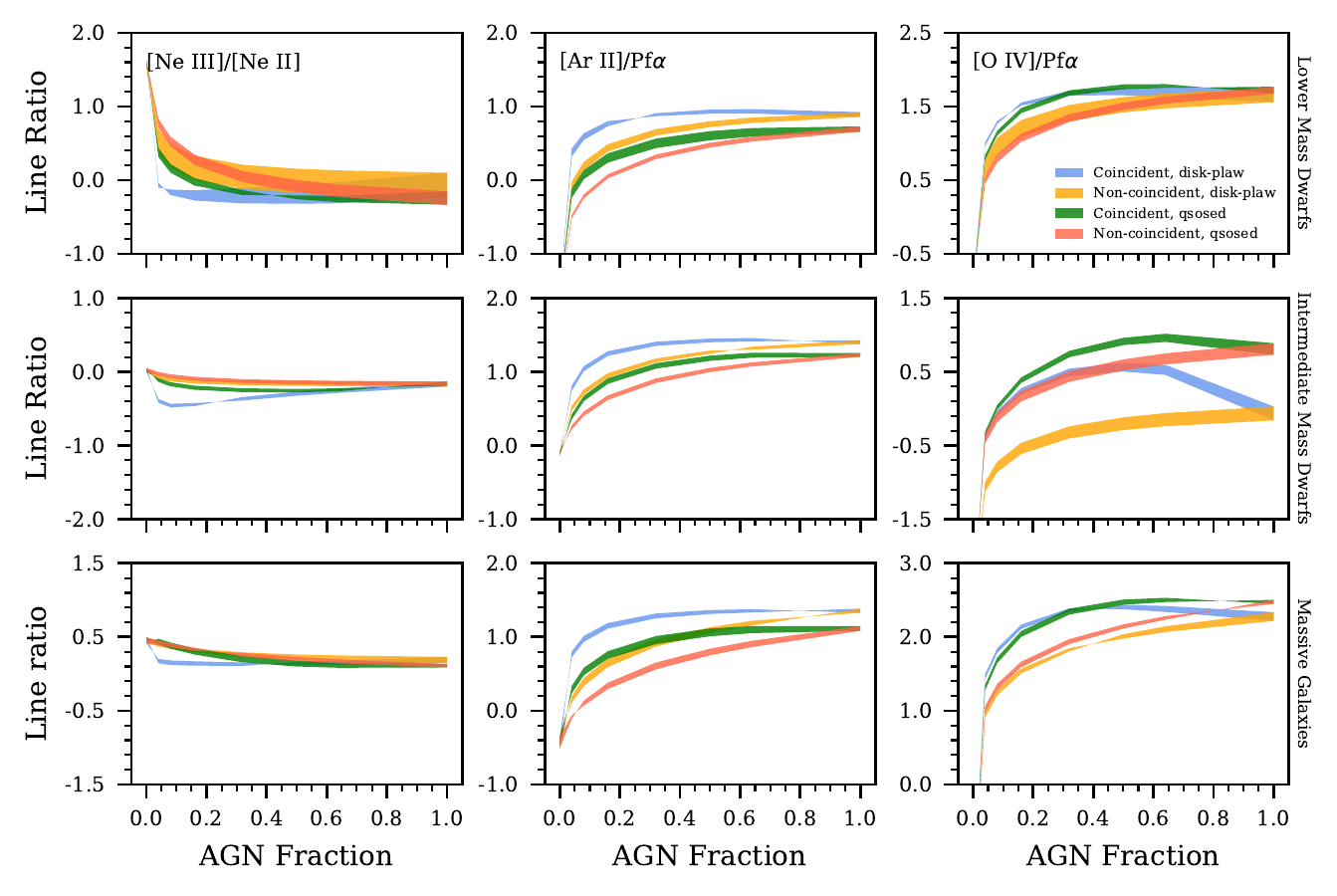}
\caption{AGN diagnostics with the JWST MIRI instrument in the same format as Fig. \ref{fig:AGN-frac-optical} assuming $M_{\mathrm{BH}} = 10^{3}~M_{\odot}$.
\label{fig:AGN-frac-IR}}
\end{figure*}

\begin{figure*}[t!]
\includegraphics[width=2.15\columnwidth]{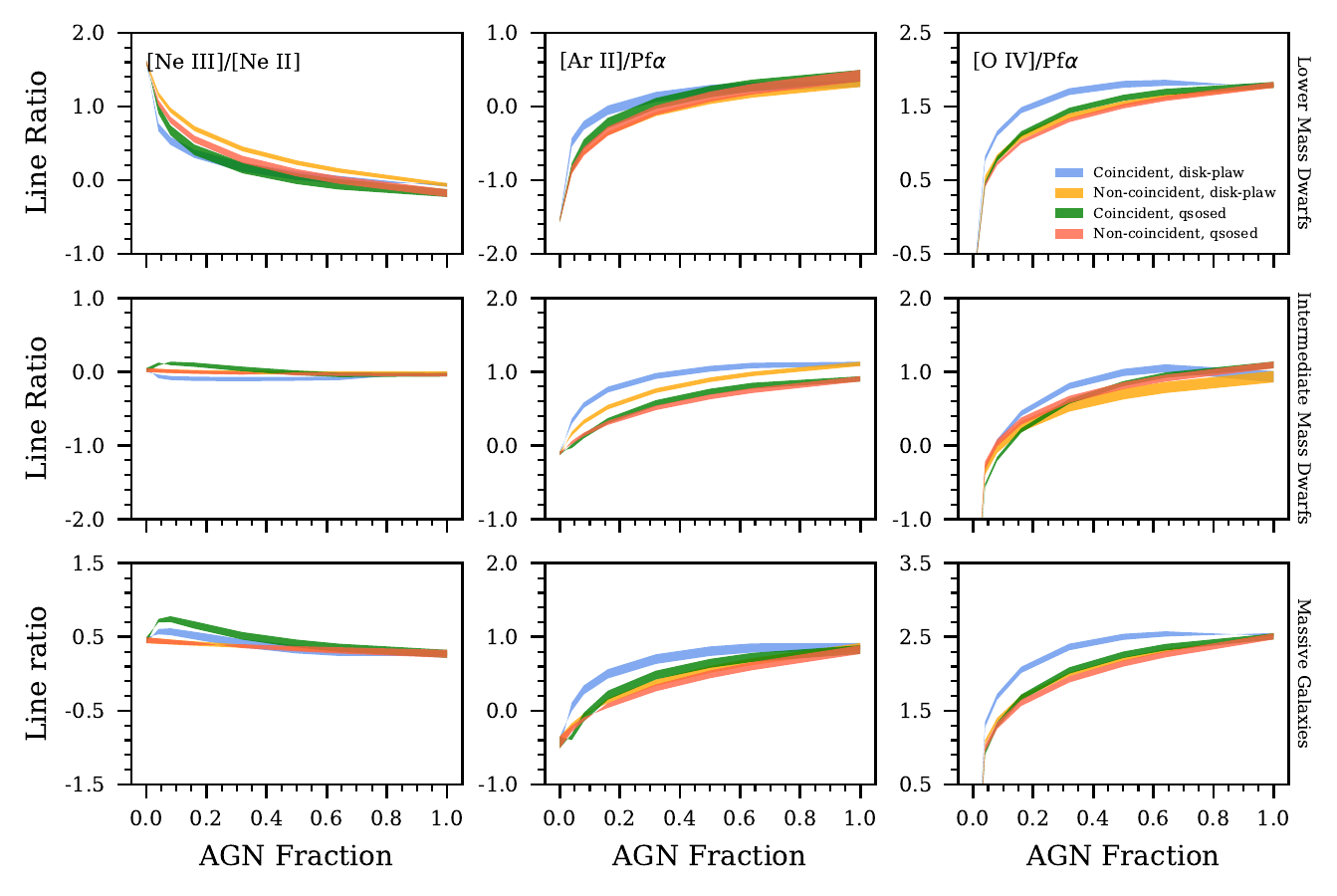}
\caption{AGN diagnostics with the JWST MIRI instrument in the same format as Fig. \ref{fig:AGN-frac-optical_M-5_0} assuming $M_{\mathrm{BH}} = 10^{5}~M_{\odot}$.
\label{fig:AGN-frac-IR_M-5_0}}
\end{figure*}

Most photoionization modeling assumes a ``standard'' set of gas phase depletion factors derived from a variety of sources. In reality, however, depletion factors $\delta_X$ depend on the undepleted reference abundance set being used and the strength of the depletion $F_*$. We follow a self-consistent approach where we use our unique reference abundances with the methodology outlined in \cite{Jenkins2009}. The strength of the depletion is selected so that the depletion factor for iron is $\delta_{Fe} = -1.5$ dex as in \cite{Thomas2018}. This results in $\delta_{O} = -0.11$ dex, which is less than the commonly assumed Cloudy default of $\delta_{O} = -0.22$ dex, but assists in matching optical emission line diagnostics (\citealt{Gutkin2016}) and matches the depletion derived from analyzing dust grain composition (\citealt{Peimbert2010}), X-ray spectroscopy (\citealt{Pinto2013}), and $\alpha$-element enhancement (\citealt{Amayo2021}). In reality, depletion factors should change with G/D at a fixed metallicity and as a function of metallicity (\citealt{Peimbert2010}, \citealt{DeCia2016}); however, this is rarely taken into account in photoionization modeling and beyond the scope of this work. The complete set of adopted depletion factors are listed in Table 1 and we elaborate on our methodology in Appendix \ref{sec:appendixA}. Our final gas phase abundances differ from the BPASS stellar abundances, the effects of which have been investigated in \cite{Grasha2021}, but are beyond the scope of this work. All in all, our model suite consists of $> 6.33\times 10^4$ simulations.


\section{Line Ratio Sensitivity Analysis\label{sec:sens-analysis}}

We seek to assess the sensitivity of emission line ratios to $f_{\mathrm{AGN}}$, AGN SED shape, mixing methodology, and geometry. In particular, we select emission line diagnostics that are detectable in purely star forming galaxies, enabling investigation of the connection between star formation and AGN in dwarfs, which is not possible for all emission lines. For example, the presence of [Ne~V]~14.3$\mu$m or [Ne~VI]~7.65 $\mu$m alone signals AGN activity, but our simulations show that [Ne~V] is unlikely to be detectable in most local dwarfs where $U$ and $f_{\mathrm{AGN}}$ are small, thus making it an unreliable tracer of AGN fraction for all galaxy masses.

Similarly, coronal lines from highly ionized states (e.g., Si~VI, Fe~XIII) have been used to identify AGN (\citealt{Cann2018}, \citealt{Bohn2021}, \citealt{Kimbro2021}). However, several limitations exist to this approach: (1) only about half of AGN actually show coronal line(s), regardless of instrumental line sensitivity (\citealt{Riffel2006}); (2) highly ionized states depend on physical conditions with high $U \sim -2.0$ and high $f_{\mathrm{AGN}} > 0.64$, which are not characteristic of typical dwarfs; (3) if present, the coronal line region lies between the BLR and NLR, which implies dust sublimation is important, a process that presents a problem for self-consistent photoionization modeling (\citealt{Mazzalay2010}, \citealt{Adhikari2016}); (4) coronal line emission typically originates from metals that become heavily depleted in forming dust grains (e.g., Si, Ca, Fe), and therefore small changes to $F_*$ yield large differences in these abundances (\citealt{DeCia2016}). All together, these limitations suggest that using coronal lines will be subject to strong selection bias, and therefore alone cannot provide a complete picture of IMBH activity in dwarf AGN.

With these caveats in mind, we have determined three different mass ranges to evaluate the AGN emission line diagnostics within the wavelength ranges of SDSS and JWST: lower mass dwarfs, intermediate mass dwarfs, and massive galaxies.

\begin{itemize}
    \item For lower mass dwarfs, we use the $U-Z$ correlation presented in \cite{Kashino2019}, which leads to values of log~$U \approx -2.0$, $Z/Z_{\odot} \approx 0.15$.
    \item For intermediate mass dwarfs, we analyzed SDSS strong emission line measurements (\citealt{Tremonti2004}) as processed by P21 for the z$\sim$0 dwarf dominated RESOLVE Survey (\citealt{Kannappan2008}) with the Bayesian analysis code \texttt{NebulaBayes} (\citealt{Thomas2018}), but using our simulation suite. We determine median values of log~$U \approx -3.25$, $Z/Z_{\odot} \approx 0.4$ for intermediate mass dwarfs (P21), which can be thought of as a `fiducial' set of parameters for dwarf galaxies.
    \item For massive galaxies, we adopt log~$U \approx -2.0$, $Z/Z_{\odot} \approx 1.0$ following the parameters used in (\citealt{Abel2009}) to model AGN in massive galaxies such as ULIRGs.
\end{itemize}

For the rest of \S\ref{sec:sens-analysis}, we use the stellar SED resulting from continuous SFH at 250 Myr as it represents a ``general purpose" ionizing continuum not purposefully constructed for maximizing any particular emission line ratio.


\subsection{Optical\label{optical}}

Figures \ref{fig:AGN-frac-optical} and \ref{fig:AGN-frac-optical_M-5_0} show the emergent optical emission line predictions for $M_{\mathrm{BH}} = 10^{3}~M_{\odot}$ and $M_{\mathrm{BH}} = 10^{5}~M_{\odot}$, respectively. In each figure, we display [O~III]/H$\beta$, [S~II]/H$\alpha$, [O~I]/H$\alpha$, three common emission line ratios used in optical AGN diagnostics (columns), as a function of AGN fraction for three galaxy mass regimes (rows). The thickness of each line denotes the uncertainty due to differences in assuming a spherical or plane parallel geometry, while the color indicates the AGN SED and mixing methodology being used. Other diagnostics are presented in Appendix \ref{sec:appendixB}, such as the metallicity sensitive [N~II]/H$\alpha$ ratio.

A few major trends are apparent in Figures \ref{fig:AGN-frac-optical} and \ref{fig:AGN-frac-optical_M-5_0}. First, the effect that physical uncertainties have on emission line ratios is exacerbated for lower $M_{\mathrm{BH}}$. Second, the greatest variation in emission line ratios due to $f_{\mathrm{AGN}}$ occurs in the range $0.0 \leq f_{\mathrm{AGN}} \leq 0.32$. The diagnostics [O~III]/H$\beta$ and [S~II]/H$\alpha$ are generally poorer tracers of $f_{\mathrm{AGN}}$. Lastly, in the dwarf galaxy mass regimes, diagnostics show less dependence on geometry, in stark contrast to massive galaxies where the diagnostics can show up to $\sim$1.0 dex variation for $M_{\mathrm{BH}} = 10^3 M_{\odot}$.

The photoionization cross section of hydrogen quickly decreases as photon energy increases (\citealt{Osterbrock2006}). As a result, X-rays due to an AGN penetrate into neutral gas, which causes collisions that excite neutral species. Since [O~I] $\lambda$6300 partially originates from neutral gas, [O~I]/H$\alpha$ traces $f_{\mathrm{AGN}}$ rather well in dwarf galaxy mass regimes in contrast to the other optical diagnostics. This makes it a reliable choice for detecting AGN activity in dwarfs and constraining $f_{\mathrm{AGN}}$. Non-ionizing photons can also pump electrons into high energy excited states, which can cause [O~I] $\lambda$6300 emission when the electrons cascade to lower energy levels. This continuum fluorescence process is rather inefficient for O~I (\citealt{Bautista1999}), leaving collision excitation as the dominant process, and thus preserving the utility of [O~I] $\lambda$6300 as an AGN diagnostic.

The AGN SED and mixing methodology can create substantial variation in all diagnostics, although this is mitigated for emission line ratios created under certain physical conditions. For example, the middle rows of Figures \ref{fig:AGN-frac-optical} and \ref{fig:AGN-frac-optical_M-5_0} show that considering $f_{\mathrm{AGN}} \geq 0.32$, the emission line ratios converge upon [O~III]/H$\beta \approx -0.3$ to $0.7$ and [O~I]/H$\alpha \approx -1.2$ to $-0.2$, generally resulting in a LINER classification (\citealt{Kewley2006}). This result, combined with a star-forming galaxy classification from the metallicity sensitive BPT diagram, corroborates that dwarf AGN have inconsistent optical classifications (\citealt{Reines2020}, P21) for black hole masses in the range $10^3 - 10^5 ~ M_{\odot}$.

Our results also stand in contrast to other work (\citealt{Cann2019}), suggesting that AGN host galaxies with $\sim 10^3 M_{\odot}$ BHs have such low [O~III]/H$\beta \approx -1.5$ that [O~III] emission might be undetectable \footnote{Private communication with Jenna Cann, Shobita Satyapal, and Nick Abel suggests that possible physical explanations could include differences in assumed AGN SED shape, in choice of elemental abundances, or in geometry and mixing, as diagrammed in Figure \ref{fig:geometry}.}. Figure \ref{fig:AGN-frac-optical} clearly shows that [O~III]/H$\beta$ is highly dependent on the physical uncertainties presented in Figure \ref{fig:geometry} and the galaxy mass regime of interest. RESOLVE dwarf galaxies are ubiquitously star-forming with $f_{\mathrm{AGN}} < 0.6$ and are the most likely hosts for IMBHs. For the intermediate mass dwarf galaxy regime, we see in the majority of cases that [O~III]/H$\beta \geq 0.0$, i.e. [O~III] as strong as H$\beta$.

\subsection{Mid-IR\label{mid-IR}}

The most valuable AGN diagnostics for JWST come from the MIRI instrument (4.9 - 28.3 $\mu$m), as opposed to NIRspec (0.6 - 5.3 $\mu$m), based on the availability of either emission lines originating from the same element with different ionization potentials, or high ionization lines along with a recombination line for comparison. Although we are focused on detecting active IMBHs in relatively local galaxies, it is worth noting that galaxies at $z>0.3$ will have optical diagnostics that fall within NIRspec, thereby providing additional constraints. The following lines show diagnostic value for AGN activity and are detectable until $z \sim 0.1$ assuming the MIRI line sensitivity of $3.3\times 10^{-17}$~erg/s/cm$^{-2}$: Pf$\alpha$, [Ne~II] 12.8$\mu$m, [Ne~III] 15.6$\mu$m, [Ar~II] 6.98$\mu$m, [Ar~III] 8.99$\mu$m, [Ar~V] 13.1$\mu$m, [S~III] 18.7$\mu$m, [S~IV] 10.5$\mu$m, and [O~IV] 25.9$\mu$m. This result is independent of geometry, mixing methodology, or AGN SED shape. 

Figures \ref{fig:AGN-frac-IR} and \ref{fig:AGN-frac-IR_M-5_0} displays tracers of AGN activity observable with JWST, in the same format as Figures \ref{fig:AGN-frac-optical} and \ref{fig:AGN-frac-optical_M-5_0}. As in the optical, changes to the line ratios become less pronounced as AGN fraction increases; in contrast to the optical, IR line ratios generally show less sensitivity to geometry. \cite{Weaver2010} suggests that [Ne~III]/[Ne~II] and [O~IV]/[Ne~III] have the ability to effectively separate excitation mechanisms. The first column in both figures indicates that [Ne~III]/[Ne~II] remains constant with AGN fraction for intermediate mass dwarfs (second row) and massive galaxies (third row). Therefore, [Ne~III]/[Ne~II] is not a robust indicator of AGN activity. It does, however, remain an effective $f_\mathrm{AGN}$ diagnostic for the low-mass dwarf galaxies, with minor sensitivity to the mixing methodology and SED selection. The second column shows that [Ar~II]/Pf$\alpha$ has strong diagnostic potential for all galaxy masses, especially for $M_{\mathrm{BH}} = 10^5 M_{\odot}$ (Figure \ref{fig:AGN-frac-IR_M-5_0}). As $M_{\mathrm{BH}}$ decreases, the [Ar~II]/Pf$\alpha$ still effectively traces $f_\mathrm{AGN}$, but the uncertainty of the AGN SED and mixing methodology introduces greater spread in the emission line predictions.

The last column in Figures \ref{fig:AGN-frac-IR} and \ref{fig:AGN-frac-IR_M-5_0} shows that [O~IV]/Pf$\alpha$ is also an effective $f_\mathrm{AGN}$ diagnostic. At $M_{\mathrm{BH}} = 10^5 M_{\odot}$, [O~IV]/Pf$\alpha$ traces $f_\mathrm{AGN}$ quite well, while remaining relatively insensitive to the shape of the AGN SED and the uncertainties given in Figure \ref{fig:geometry}. [O~IV]/Pf$\alpha$ remains well-behaved for $M_{\mathrm{BH}} = 10^3 M_{\odot}$ except for intermediate mass dwarfs where the disk-plaw, non-coincident mixing case creates a $\sim$1.0 dex spread in values relative to the other cases. This spread is due to the combined effect of a low $U$ and harder accretion disk reducing the number O$^{++}$ ionizing photons.

\textit{In general, line ratios featuring [O~IV] and [Ar~II] show less sensitivity to geometry than optical diagnostics, display more variation with AGN fraction than other mid-IR line ratios, and retain their diagnostic potential over a wider range of galaxy masses and BH masses.} The mid-IR diagnostics provide greater constraints on $f_{\mathrm{AGN}}$ for lower levels of AGN activity expected for dwarf AGN with IMBHs, compared to optical diagnostics which are more sensitive to higher levels of AGN activity. These results highlight the benefit of using multiple line ratios to constrain $f_{\mathrm{AGN}}$ with diagnostic diagrams.

\begin{figure*}[t!]
\includegraphics[width=2.15\columnwidth]{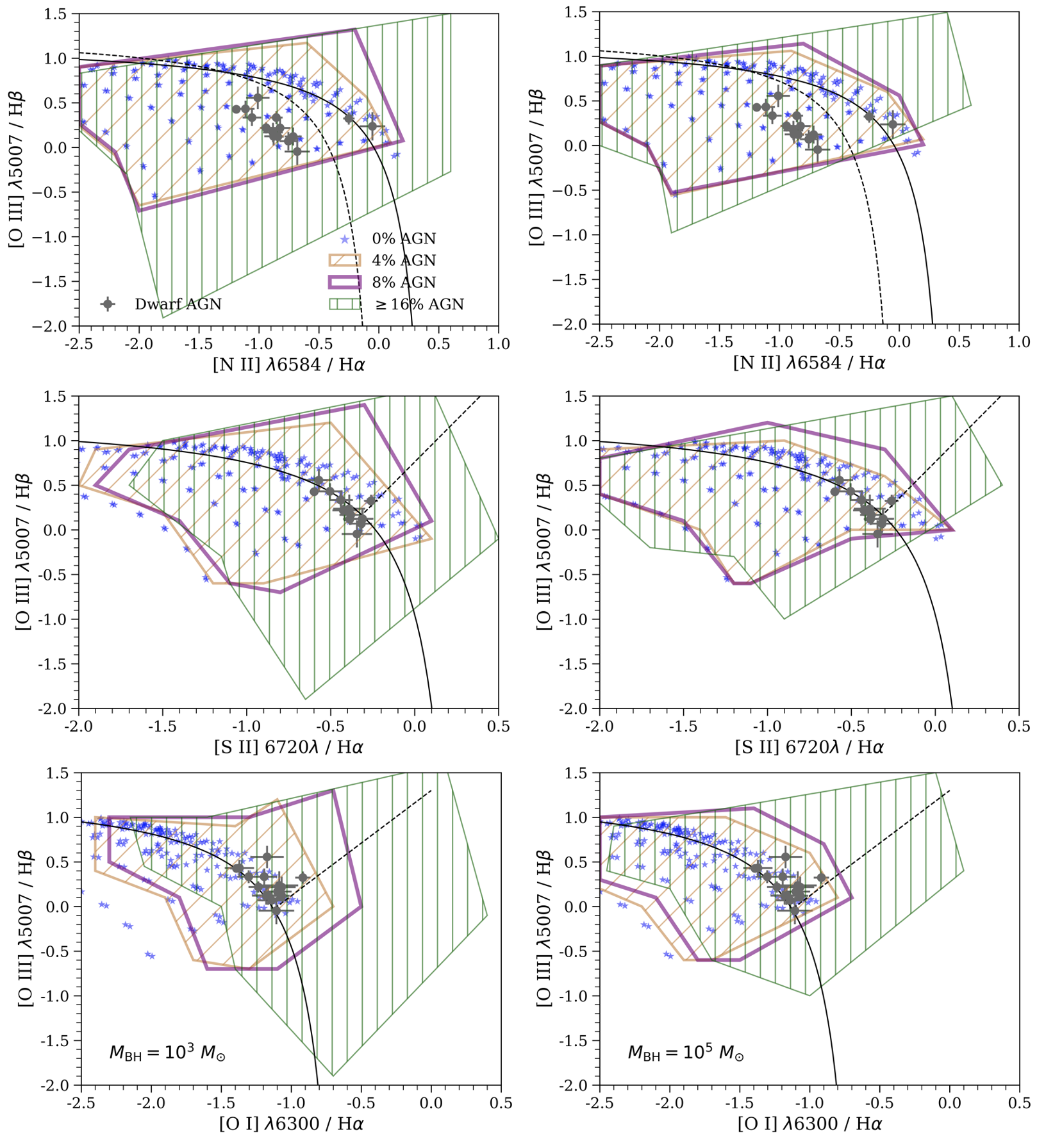}
\caption{Optical diagnostic diagrams that separate AGN activity from starburst activity assuming $M_{\mathrm{BH}} = 10^3 M_{\odot}$ (left column) and $M_{\mathrm{BH}} = 10^5 M_{\odot}$ (right column). The selected age of the binary stellar population maximizes the contribution of WR stars and thereby sets a lower limit for AGN activity. Pure starburst models (0\% AGN) are displayed as blue stars. The models for a given $f_{\mathrm{AGN}}$ approximately span the area of each hatched shape. The demarcations in each diagram are taken from \cite{Kewley2001}, \cite{Kauffmann2003}, and \cite{Kewley2006}. The majority of observed dwarf AGN lie in the star forming wing of the BPT diagram (top panels), which models with and without AGN can reproduce.
\\
\\
\label{fig:optical-ddiag}}
\end{figure*}

\section{Diagnostics Diagrams}

In light of the wide range of uncertainties previously mentioned, we seek to find diagnostic diagrams suitable for detecting AGN excitation, and ideally measuring $f_{\mathrm{AGN}}$, for a given galaxy. To compare our models to galaxies with spectroscopic observations, we select a subset of models ($-3.5 \leq \mathrm{log}~U \leq -1.5$) with an instantaneous SFH, since the harder stellar continua provide a limiting case for AGN activity.

\subsection{Optical \label{subsec:ddiag_optical}}

We use a 20 Myr SSP for the optical diagnostics since this SFH maximizes the [O~III]/H$\beta$ line ratio used in many diagnostic diagrams that separate AGN and star forming galaxies. For the observational sample, we choose dwarf galaxies in the MPA-JHU sample from P21. Following P21, we classify dwarf AGN as any dwarf galaxies not classified as definitely star forming (e.g., traditional AGN, LINERs, etc.) The three diagrams from \cite{V&O87} are by far the most commonly used optical diagnostics for separating AGN and star forming galaxies, so we focus our analysis on them.

Figure \ref{fig:optical-ddiag} displays these diagrams for $M_{\mathrm{BH}} = 10^3 M_{\odot}$ (left column) and $M_{\mathrm{BH}} = 10^5 M_{\odot}$ (right column). The models for $f_{\mathrm{AGN}} =$ 0.04, 0.08, and $\geq$0.16 approximately span the area of the three hatched shapes (tan, purple, and green, respectively). For models with $f_{\mathrm{AGN}} \geq 0.16$ and a $10^3 M_{\odot}$ black hole, there is a wide range of possible values for [O~III]/H$\beta$, highly dependent on configurations present in Figure \ref{fig:geometry}. This large spread in values relates to the current difficulty of detecting IMBHs at this mass, but also provides promise that optical detection is indeed possible. 

Figure \ref{fig:optical-ddiag} shows that models in the range $0 \leq f_{\mathrm{AGN}} \leq 0.16$ occupy similar regions of each diagram. In particular, the BPT diagram (top row) registers observed dwarf AGN as consistent with pure star formation. Conversely, theoretical models with pure star formation can cross over into the AGN region of the diagram. Despite the differences in photoionization modeling, this is also seen in \cite{Xiao2018} on account of using the BPASS SEDs that feature substantially harder continua than many other SPS codes.

The [S~II]/H$\alpha$ diagram indicates that pure star formation can reproduce the line ratios of all the dwarf AGN shown, but unlike for the BPT diagram, for the [S~II]/H$\alpha$ diagram this result only occurs with the particularly hard stellar SED that we have selected. Thus, knowing the SFHs for a given sample of dwarfs could provide justification for using [S~II]/H$\alpha$ to identify AGN. In contrast, the BPT diagram is more generally a poor diagnostic for identifying dwarf AGN, regardless of SFH, as discussed in more detail in P21.

The last optical diagnostic diagram shows that some dwarf AGN require models with an AGN component to reproduce the observed [O~I]/H$\alpha$, making this diagram more reliable in separating star formation and AGN in dwarfs (\citealt{Reines2020}, P21). It is worth noting that more dwarf AGN likely require an AGN component than shown here, on account of the hard stellar SED used in this analysis. Additionally, for [O~I]/H$\alpha$ there is some modest separation between models with different $f_{\mathrm{AGN}}$, especially with $M_{\mathrm{BH}} = 10^3 M_{\odot}$, enabling a better assessment of AGN activity as also shown in Figure \ref{fig:AGN-frac-optical}.

\subsection{Mid-IR \label{subsec:ddiag_IR}}

For our observational sample, we choose dwarf SF (\citealt{Cormier2015}) and dwarf AGN (\citealt{Hood2017}) observations requiring that the AGN satisfy [Ne~V]/[Ne~II]$>$0.1 (\citealt{Inami2013}). It is important to note that Pf$\alpha$, featured in Figure \ref{fig:AGN-frac-IR}, is not available in statistical samples due to the low sensitivity and resolution of Spitzer and ISO. Similarly, since [Ar~II] $6.98\mu$m is unavailable for samples including dwarfs, we use massive AGN (\citealt{Sturm2002}) for diagnostic diagrams with this line.

\begin{figure*}
\includegraphics[width=2.1\columnwidth]{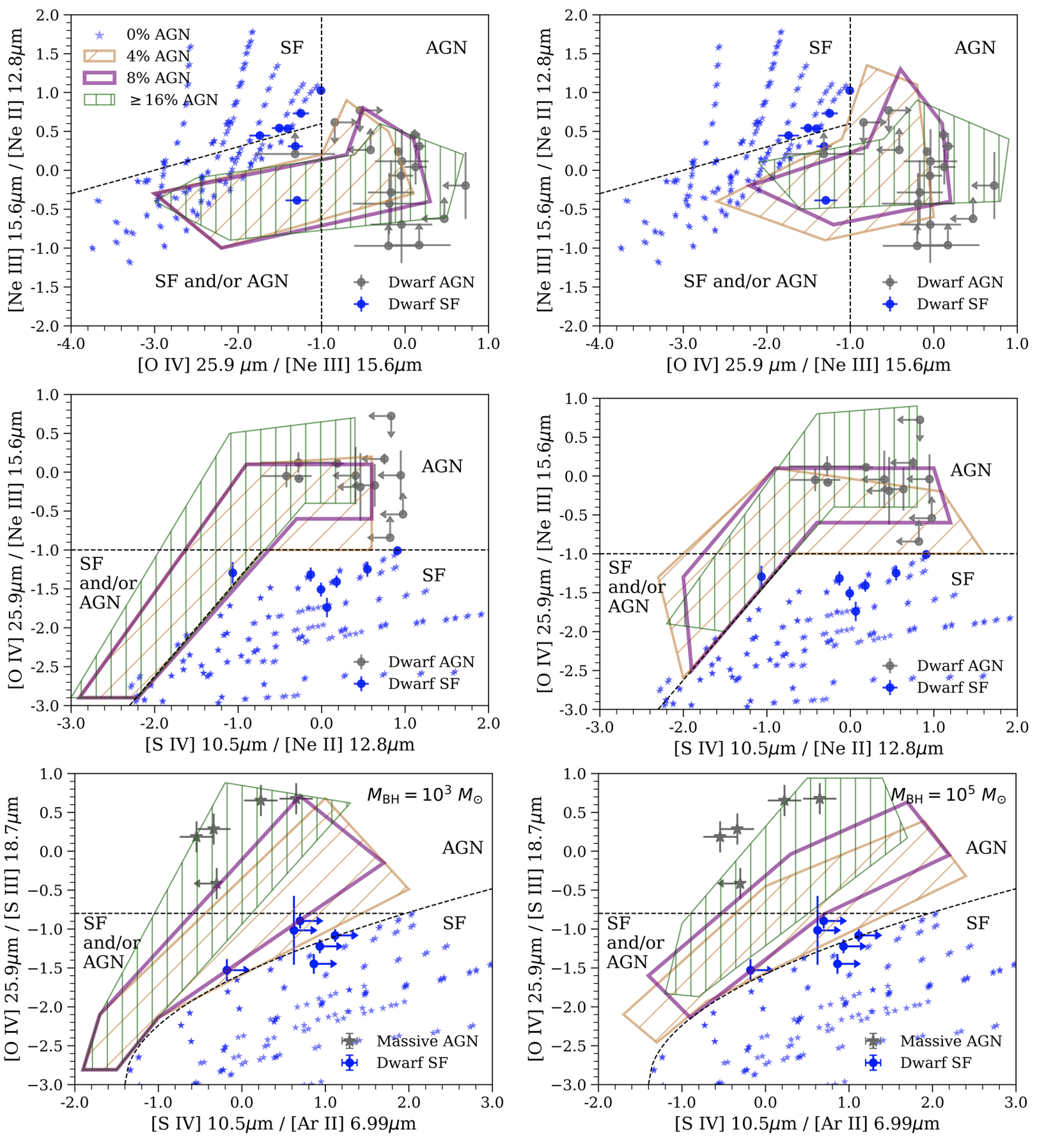}
\caption{Mid-IR diagnostic diagrams that separate AGN activity from pure starburst activity in the same format as Figure~\ref{fig:optical-ddiag}. Pure starburst models (0\% AGN) are displayed as blue stars. The models for a given $f_{\mathrm{AGN}}$ approximately span the area of each hatched shape. The SF, SF and/or AGN, and AGN regions represent pure starburst galaxies, a mix of pure starburst galaxies and AGN, and only AGN, respectively (eqns 6, 7, 8). JWST will enable emission lines normally missing from Spitzer spectra to serve as additional constraints, like [Ar~II] 6.99 $\mu$m (left panels).
\label{fig:IR-ddiag}}
\end{figure*}

We have tested all possible emission line ratio combinations from the emission lines listed in \S3.1.2. Figure~\ref{fig:IR-ddiag} shows the results of this analysis in three diagnostic diagrams featuring emission line ratios also presented in \cite{Weaver2010}, \cite{Inami2013}, and \cite{Hao2009}.

Our new demarcations separate three distinct regions: the SF region contains only pure starburst galaxies; the SF and/or AGN region contains a mixture of pure starbursts and AGN; the AGN region contains only AGN. The divisions between SF and SF and/or AGN regions are given by,

\begin{equation}
\textrm{log [Ne III]/[Ne II]} = 0.3 ~ \textrm{log [O IV]/[Ne III]} - 0.9
\label{eqn:weaver-demarc}
\end{equation}

\begin{equation}
\resizebox{.9\hsize}{!}{\textrm{log[O IV]/[Ne III]} = 1.25~\textrm{log [S IV]/[Ne II]} - 0.125}
\label{eqn:inami-demarc}
\end{equation}

\begin{equation}
\resizebox{.9\hsize}{!}{\textrm{log [O IV]/[S III]} = 1.2~(\textrm{log [S IV]/[Ar II]}+1.4)$^{\frac{1}{2}}$ - 3.0}
\label{eqn:richardson-demarc}
\end{equation}

\noindent while the AGN region is formed by [O~IV]/[Ne~III]$~= -1$ and [O~IV]/[S~III]$~= -0.8$.

The first diagram (top panels) has frequently been used to separate starbursts and AGN into quadrants according to [O~IV]/[Ne~III] $=0$ (\citealt{Weaver2010}), but this cutoff excludes dwarf AGN with [O~IV]/[Ne~III]~$>-1$, which our new demarcations recover. Our models indicate that the observed dwarf AGN in this region have at least a 4-16\% AGN contribution. The second diagram (center panels) replaces the abscissa with [S~IV]/[Ne~II] (\citealt{Inami2013}), resulting in a cleaner separation of starbursts and AGN at low ionization. The last diagram (bottom panels) uses [O~IV]/[S~III] in the ordinate (\citealt{Hao2009}) and [S~IV]/[Ar~II] in the abscissa, which creates even more separation between starbursts and AGN, and also between $f_{\mathrm{AGN}}$ values. As mentioned above, the [Ar~II] observations are poor since Spitzer was only capable of detecting wavelengths $<$10 microns in low resolution mode ($R \sim 130$). In contrast, MIRI features the highest resolution ($R \sim 3100$) and line sensitivity available at these wavelengths. Thus, the right panels in Figure \ref{fig:IR-ddiag} present a promising new diagnostic for JWST observations to constrain AGN activity.

\section{Discussion and Conclusions}

In this paper, we have explored several key uncertainties that will be critical to assess in the upcoming era of detecting IMBHs: the shape of the AGN SED, the surrounding gas cloud geometry, and the manner of mixing radiation from AGN and stars. The diversity of models for the AGN SED in Figure \ref{fig:seds} emphasizes the importance of next generation X-ray facilities such as Lynx and Athena that will enable proper constraints on low luminosity sources. We show that all of the listed uncertainties have profound effects on emission line diagnostics in the optical and mid-IR. This has implications for statistical samples of galaxies that are analyzed with photoionization models that implicitly assume a particular geometrical configuration or mixing methodology in Figure \ref{fig:geometry}. Such assumptions can lead to powerful selection effects (\citealt{Ferguson1997}, \citealt{Meskhidze17}). These assumptions could limit the applicability of simple relationships that scale physical properties (e.g., $f_{\mathrm{AGN}}$, $M_{\mathrm{BH}}$) with emission lines ratios for large samples of galaxies.

Even though degeneracies abound in the optical, we find AGN diagnostics remain detectable in most situations across a range of galaxy masses. This result offers promise that surveys like SDSS have already picked up signatures of AGN with $M_{\mathrm{BH}} = 10^3 ~ M_{\odot}$. Unfortunately, the gold standard for optical AGN classification, the BPT diagram, poorly traces AGN activity in dwarf galaxies. As we argue in P21, this is due to [N~II]/H$\alpha$ being metallicity sensitive and $f_{\mathrm{AGN}}$ insensitive. It also reflects the sensitivity of [O~III]/H$\beta$ to the uncertainties explored in this paper. Actually, as seen in \citealt{Reines2020} and P21, the often forgotten [O~I]/H$\alpha$ diagnostic is a stronger metric for an active AGN in dwarfs given its sensitivity to $f_{\mathrm{AGN}}$ and relative insensitivity to physical uncertainties. It therefore serves as the best optical emission line ratio for finding dwarf AGN, but the relative weakness of the [O~I] line does limit the sample that can be tested (P21).

Relatively high-mass dwarfs ($M_* \lesssim 10^{9.3-9.5} M_{\odot}$) classified as LINERs present an opportunity to find the coveted $10^3 ~ M_{\odot}$ black holes, which would fill in a major gap in IMBH empirical relationships and help constrain the occupation fraction in dwarfs. If AGN in this BH mass regime are confirmed by other methods (e.g., variability, line broadening, X-rays), but still fail to show any optical signatures, this non-detection would suggest that the column density of obscuring gas and the covering factor around the source could be factors in the elusive nature of these AGN. The column densities and covering factors of dwarf AGN remain relatively unexplored. Purely star forming dwarf galaxies show a decrease in covering factor as metallicity decreases, albeit with a large scatter in the relation (\citealt{Cormier2019}). Conversely, massive galaxies show a decrease in obscured AGN (large covering factor) with increasing luminosity (\citealt{Sazonov2015}, \citealt{Georgakakis2017}), and therefore increasing metallicity (\citealt{Lamareille2004}). It is possible that dwarf AGN could provide the missing link between these two opposing trends if low mass IMBHs continue to remain elusive in the optical.

In the mid-IR, we have shown that [Ar~II] 6.98$\mu$m / Pf$\alpha$ and [O~IV] 25.9$\mu$m/Pf$\alpha$ together have the potential to constrain AGN activity over a range of galaxy masses while minimizing sensitivity to physical uncertainties (Figure~\ref{fig:AGN-frac-IR}). We have revised the demarcations on the [Ne~III]/[Ne~II] vs. [O~IV]/[Ne~III] diagnostic diagram to include dwarf AGN and presented two new diagnostic diagrams based on the [O~IV] and [Ar~II] emission lines (Figure \ref{fig:IR-ddiag}). These new diagrams are capable of separating starburst activity from AGN activity down to a 4\% AGN fraction in dwarfs, which are ubiquitously star forming. Unlike the BPT diagram, these proposed diagrams maintain their diagnostic value over a wide range of ionization and metallicity.

We emphasize that the provided demarcations serve as theoretical boundaries for classifying the excitation mechanism in dwarf galaxies. Observations of optically classified starbursts can occasionally cross over into the AGN region of these diagrams due to strong mid-IR [O~IV] emission (e.g., IZw18, \citealt{Lebouteiller2017}), although soft X-ray and hard X-ray emission is typically detected in these galaxies.

Current SPS models are unable to produce enough photons $>$54~eV to account for this crossover. High mass X-ray binaries (HMXBs) can generate the hard photons needed to create AGN-like emission line ratios assuming a model with high $U$, high $L_X/\mathrm{SFR}$, and a sufficiently hard SED (\citealt{Simmonds2021}). It is unlikely that HMXBs can uniformly account for dwarfs exhibiting emission line ratios characteristic of AGN given that the shape of the HMXB SED remains highly uncertain and most dwarf AGN do not present these extreme conditions (P21). Additionally, a more generic treatment of the HMXB SED has shown that HMXBs are an inefficient means of producing photons $>$54~eV (\citealt{Senchyna2020}). Therefore, unless more realistic treatment of the WR phase alters the result of our predictions in the future, our current simulations suggest these galaxies have active AGN. The SF and/or AGN region in each diagram can contain optically classified starbursts (see Figure \ref{fig:IR-ddiag}), and this result is reproduced by a photoionization model with multiple gas clouds (\citealt{Melendez2014}, \citealt{Richardson2016}). IFU observations should help clarify whether such a model needs to be invoked.

To fully address modeling degeneracies in both the optical and mid-IR, Bayesian analysis will be a valuable tool for extracting meaningful properties over a range of conditions in future studies. Our simulation suite incorporates a finer spacing of metallicities, ideal for accurate Bayesian analysis at $< 0.4~Z_{\odot}$ (\citealt{Richardson2019}), than other SPS models (e.g., Starburst99), while robustly accounting for D/G and elemental depletion. In addition, the mixing methodologies and geometries unique to our models could be used in a multi-component ISM analysis, similar to work on the multi-phase ISM in dwarfs (\citealt{Cormier2019}). Together with the AGN diagnostics and excitation diagrams presented in this paper, such future applications of our models will prove useful for addressing more complex topologies associated with mixed AGN/SF excitation in dwarfs in the era of JWST.


\acknowledgments

CR gratefully acknowledges the support of the Elon University FR\&D committee and the Extreme Science and Engineering Discovery Environment (XSEDE), which is supported by National Science Foundation grant number ACI-1548562. This work used the XSEDE resource Comet at the San Diego Supercomputing Center through allocation TG-AST140040. MP gratefully acknowledges the support of the 2020 Hamilton Award from the UNC Department of Physics and Astronomy. SK and MP acknowledge support from NSF AST-2007351. JB acknowledges grant support from NSF AST-1812642 and CUNY JFRASE. We thank Nick Abel, Jenna Cann, Chris Done, Gary Ferland, Ed Jenkins, and Shobita Satyapal for helpful discussions that improved the quality of this paper.

%

\vspace{5mm}


\software{Cloudy \citep{Ferland2017},
          NebulaBayes \citep{Thomas2018}.
          }

\clearpage
\bibliography{mixing-JWST}{}
\bibliographystyle{aasjournal}






\appendix

\section{Depletion Factors Due to Grains \label{sec:appendixA}}

We use the methodology presented in \cite{Jenkins2009} (hereafter J09) to develop a set of depletion factors that account for gas phase elements condensing to form dust grains. This depletion depends on comparing a reference abundance set to abundances derived from observations along a particular line of sight. The difference provides the depletion factor, $\delta_X$,

\begin{equation}
\delta_X = \mathrm{log} \left( \frac{X}{\mathrm{H}} \right)_{obs} - \mathrm{log}\left( \frac{X}{\mathrm{H}} \right)_{ref}
\end{equation}

\noindent Despite the dependence on a reference set of abundances, most photoionization modeling does not properly take this factor into account as shown below. Different lines of sight yield different depletion factors for a given element, allowing one to define a depletion strength, $F_*$, that accounts for this variation. A linear fit to the logarithm of the depletion factor gives the following form,

\begin{equation}
\delta_X = B_X + A_X (F_*-z_X)
\end{equation}

\noindent where $A_X$ is the slope, $B_X$ is the vertical offset, and $z_X$ accounts for the errors in the observations. We assume that the non-refractory elements He, Ne, S, and Ar do not become depleted in the ISM. However, sulphur depletion remains a subject of debate (\citealt{Gry2017}, \citealt{Laas2019}, \citealt{Goicoechea2021}, \citealt{Hily-Blant2021}). 

We use the fits for the Galaxy provided in J09 for C, N, O, Mg, Si, P, Cl, Ti, Cr, Mn, Fe, Ni, Cu, and Zn. We make this choice for two reasons. First, J09 provides the most complete sample, while extragalactic fits are mostly limited to heavier elements that have less impact on emission line predictions. Second, the number of observations included in the J09 analysis makes the fit more reliable.

While we assume that depletion occurs as observed in the Milky Way, deviations in the depletion patterns for heavy elements are known to be present for other galaxies. For example, Mg, Ti, and Mn show noticeable deviations in the Small Magellanic Cloud (\citealt{Jenkins2017}), and while the depletion pattern is mostly Galactic in the Large Magellanic Cloud (\citealt{Tchernyshyov2015}), Si does show a noticeable deviation (\citealt{Roman-Duval2019}).

For the rest of the elements, we have compiled a list of depletion factors towards the highly depleted star $\zeta$ Oph to serve as the value at $F_* = 1.0$ in accordance with the method used by J09. Fluorine is the only exception where we use the most depleted source in \cite{Snow2007}, since the value towards $\zeta$ Oph is unavailable. To determine the parameters $A_X$ and $B_X$ for these elements, we rely on the general trend that elements with larger condensation temperatures show greater depletions and steeper slopes. The following elements have similar depletion trends to their J09 counterparts: Li $\Leftrightarrow$ Cr; Be $\Leftrightarrow$ Mg; B $\Leftrightarrow$ Zn, F $\Leftrightarrow$ Cl; Na, K $\Leftrightarrow$ Cu, Zn; Al, Ca, V $\Leftrightarrow$ Ti; Sc $\Leftrightarrow$ P, Cu; Co $\Leftrightarrow$ Ni. From these analogies, we can assume that all of the elements in J09 have $\delta_X = 0$ at $F_* \approx -0.5$ except for $\delta_{\mathrm{Fluorine}}$, which reaches zero at $F_* = 0.2$.

Table \ref{tab:depletion_table} lists the depletion factors for all elements for $F_* = 1.0$, the linear fit parameters need to recreate the depletion pattern, and the reference for each depletion factor given. Note that depletion factors have been rescaled according to our reference abundances instead of using the abundance assumed in each cited source. Displaying the depletion factors for $F_* = 1.0$ makes it clear which elements are weakly depleted in the ISM as opposed to not depleted at all. However, for our photoionization modeling, we adjusted $F_*$ so that $\delta_{\mathrm{Fe}} = -1.5$ (see Table \ref{tab:abundance_table}), which we justified in \S \ref{sec:gas-cloud}.

The methodology we have used here is similar to what resulted in the depletion factor sets included with the photoionization code \texttt{Mappings V} (\citealt{Sutherland2017}) with two major differences. First, unlike the values included with \texttt{Mappings}, our depletion factors are self-consistently scaled  with our reference abundances. Second, we do not deplete nitrogen for any value of $F_*$ given the lack of evidence that nitrogen gets locked up in grains. This is particularly important for any analysis that strongly relies upon nitrogen emission lines for inferring metallicity values.

The depletion factors we use represent a small step toward accounting for the different depletion patterns in gas phase abundances, much in the same way \cite{Nicholls2017} accounts for nucleosynthesis patterns for elements as a function of metallicity. However, obvious limitations exist since galaxies can show more negative $F_*$ than we account for here (\citealt{Jenkins2017}).

\begin{table}
	\centering
	\caption{Depletion factors adjusted for our reference abundance set at $F_* = 1.0$ along with the parameters necessary to scale each element with $F_*$. The non-refractory elements He, Ne, S, and Ar are assumed to have no depletion.}
	\label{tab:depletion_table}
	\begin{tabular}{lccccl} 
		\hline
		X & $\delta_X (F_* = 1.0)$ & $A_X$ & $B_X$ & $z_X$ & Ref. \\
		\hline
		He & 0.0 & - & - & - & - \\
		Li & -0.827 & -0.552 & -0.28 & - & \citealt{White1986} \\
		Be & -0.432 & -0.288 & -0.14 & - & \citealt{York1982} \\
		B & -0.862 & -0.575 & -0.29 & - & \citealt{Federman1993} \\
		C & -0.176 & -0.101 & -0.16 & 0.803 & \citealt{Jenkins2009} \\
		N & 0.000 & 0.0 & 0.0 & 0.550 & \citealt{Jenkins2009} \\
		O & -0.235 & -0.225 & -0.15 & 0.598 & \citealt{Jenkins2009} \\
		F & -0.470 & -0.587 & 0.117 & - & \citealt{Snow2007} \\
		Ne & 0.0 & - & - & - & - \\
		Na & -0.850 & -0.567 & -0.283 & - & \citealt{Savage1996} \\
		Mg & -1.208 & -0.997 & -0.74 & 0.531 & \citealt{Jenkins2009} \\
		Al & -2.530 & -1.687 & -0.843 & - & \citealt{Barker1984} \\
		Si & -1.250 & -1.136 & -0.46 & 0.305 & \citealt{Jenkins2009} \\
		P & -0.520 & -0.945 & -0.04 & 0.488 & \citealt{Jenkins2009} \\
		S & 0.0 & - & - & - & -\\
		Cl & -0.720 & -1.242 & -0.23 & 0.609 & \citealt{Jenkins2009} \\
		Ar & 0.0 & - & - & - & - \\
		K & -0.970 & -0.647 & -0.323 & - & \citealt{Chaffee1982} \\
		Ca & -3.720 & -2.480 & -1.24 & - & \citealt{Crinklaw1994} \\
		Sc & -2.421 & -1.614 & -0.807 & - & \citealt{Snow1980} \\
		Ti & -3.054 & -2.048 & -1.89 & 0.43 & \citealt{Jenkins2009} \\
		V & -1.830 & -1.220 & -0.61 & - & \citealt{Savage1996} \\
		Cr & -2.175 & -1.447 & -1.41 & 0.47 & \citealt{Jenkins2009} \\
		Mn & -1.605 & -0.857 & -1.19 & 0.52 & \citealt{Jenkins2009} \\
		Fe & -2.216 & -1.285 & -1.49 & 0.437 & \citealt{Jenkins2009} \\
		Co & -2.120 & -1.413 & -0.707 & - & \citealt{Mullman1998} \\
		Ni & -2.336 & -1.49 & -1.74 & 0.599 & \citealt{Jenkins2009} \\
		Cu & -1.147 & -0.71 & -0.94 & 0.711 & \citealt{Jenkins2009} \\
		Zn & -0.410 & -0.61 & -0.14 & 0.555 & \citealt{Jenkins2009} \\
		\hline
	\end{tabular}
\end{table}

\section{Additional Line Ratio Sensitivity Diagrams \label{sec:appendixB}}

\begin{figure*}
\includegraphics[width=1.0\columnwidth]{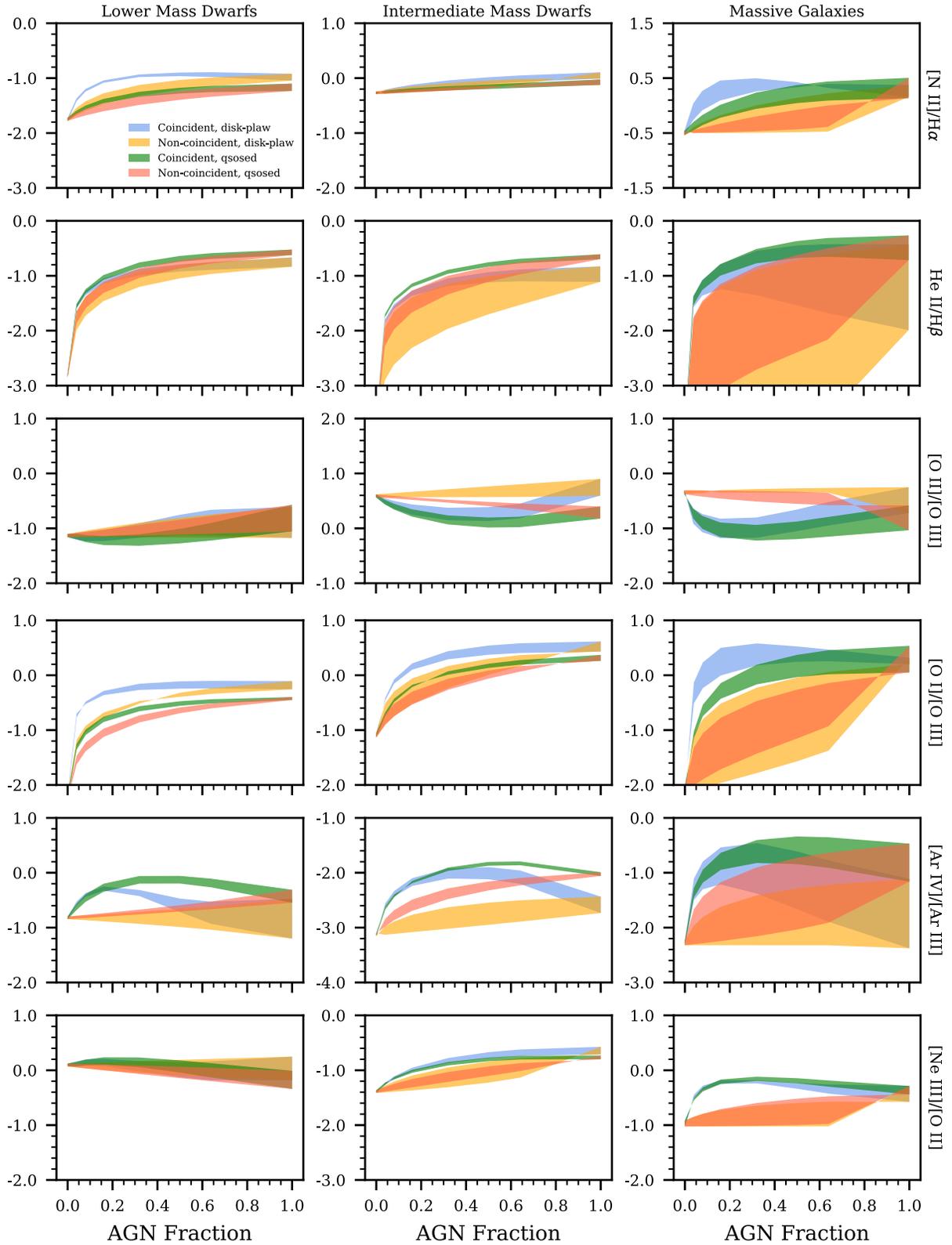}
\caption{Additional optical emission line excitation diagnostics in the same format as Figure \ref{fig:AGN-frac-optical} with $M_{\mathrm{BH}} = 10^3~M_{\odot}$.
\label{fig:appendix_optical_M-3_0}}
\end{figure*}

\begin{figure*}
\includegraphics[width=1.0\columnwidth]{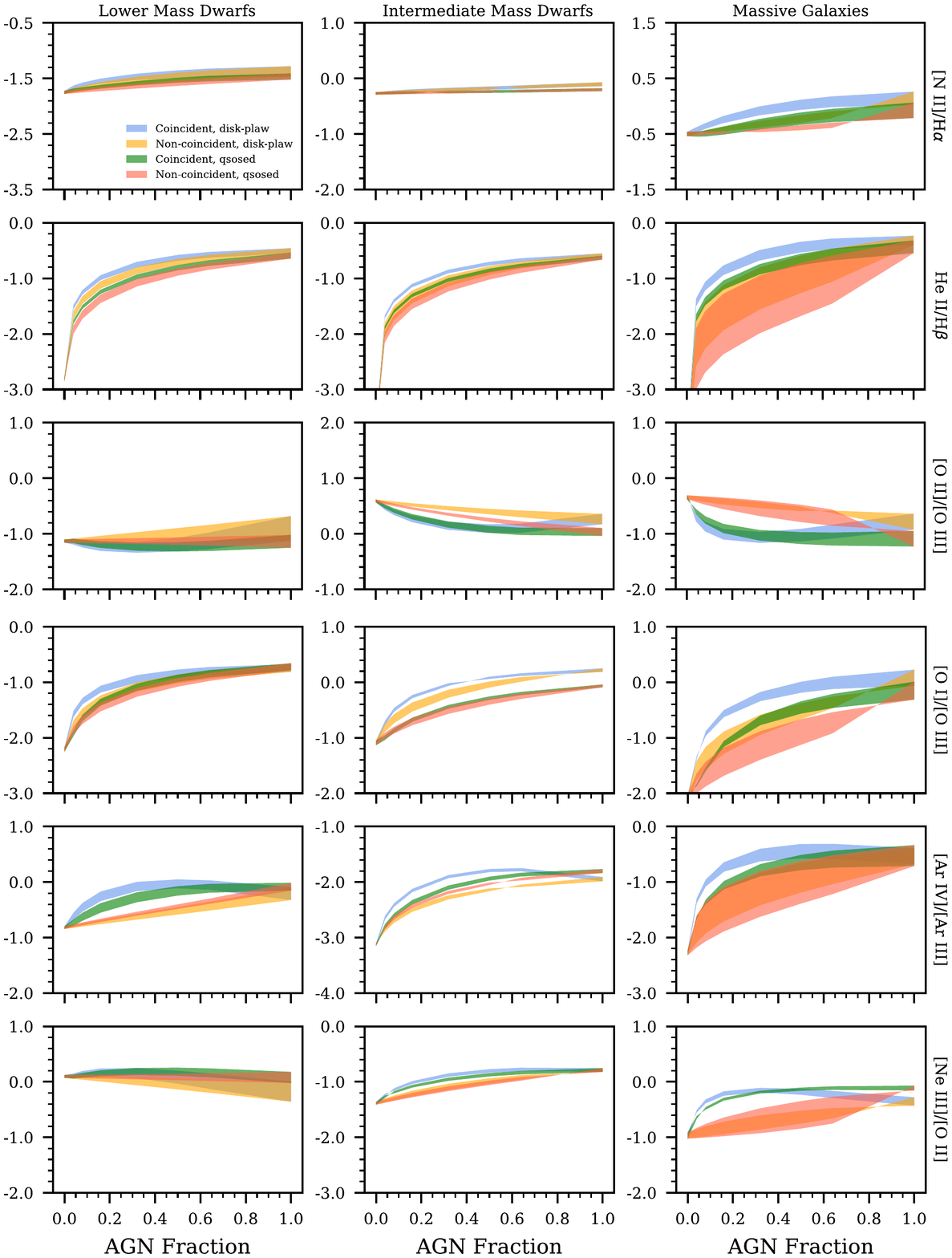}
\caption{Additional optical emission line excitation diagnostics in the same format as Figure \ref{fig:AGN-frac-optical_M-5_0} with $M_{\mathrm{BH}} = 10^5~M_{\odot}$.
\label{fig:appendix_optical_M-5_0}}
\end{figure*}

\begin{figure*}
\includegraphics[width=1.0\columnwidth]{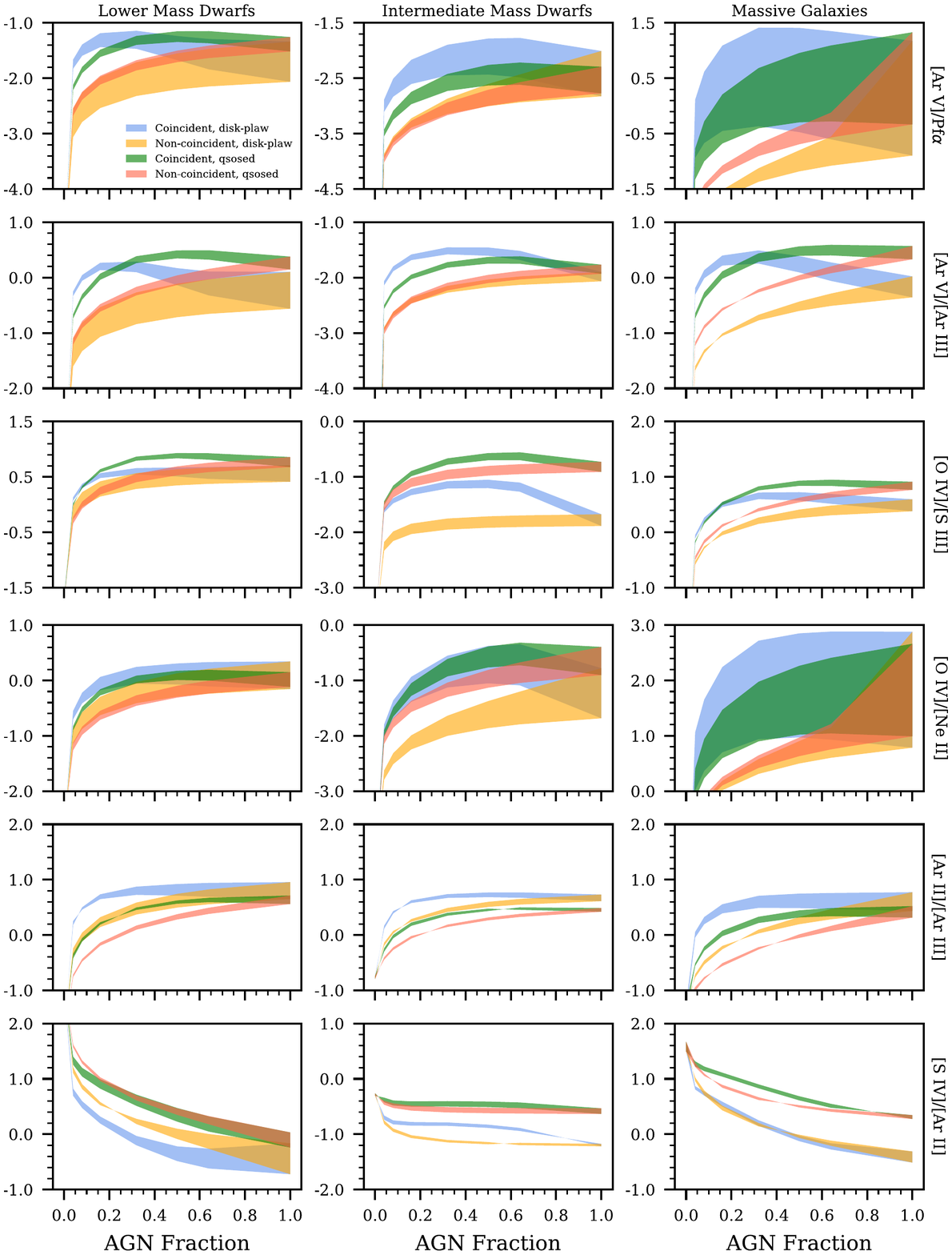}
\caption{Additional mid-IR emission line excitation diagnostics in the same format as Figure \ref{fig:AGN-frac-IR} with $M_{\mathrm{BH}} = 10^3~M_{\odot}$.
\label{fig:appendix_IR_M-3_0}}
\end{figure*}

\begin{figure*}
\includegraphics[width=1.0\columnwidth]{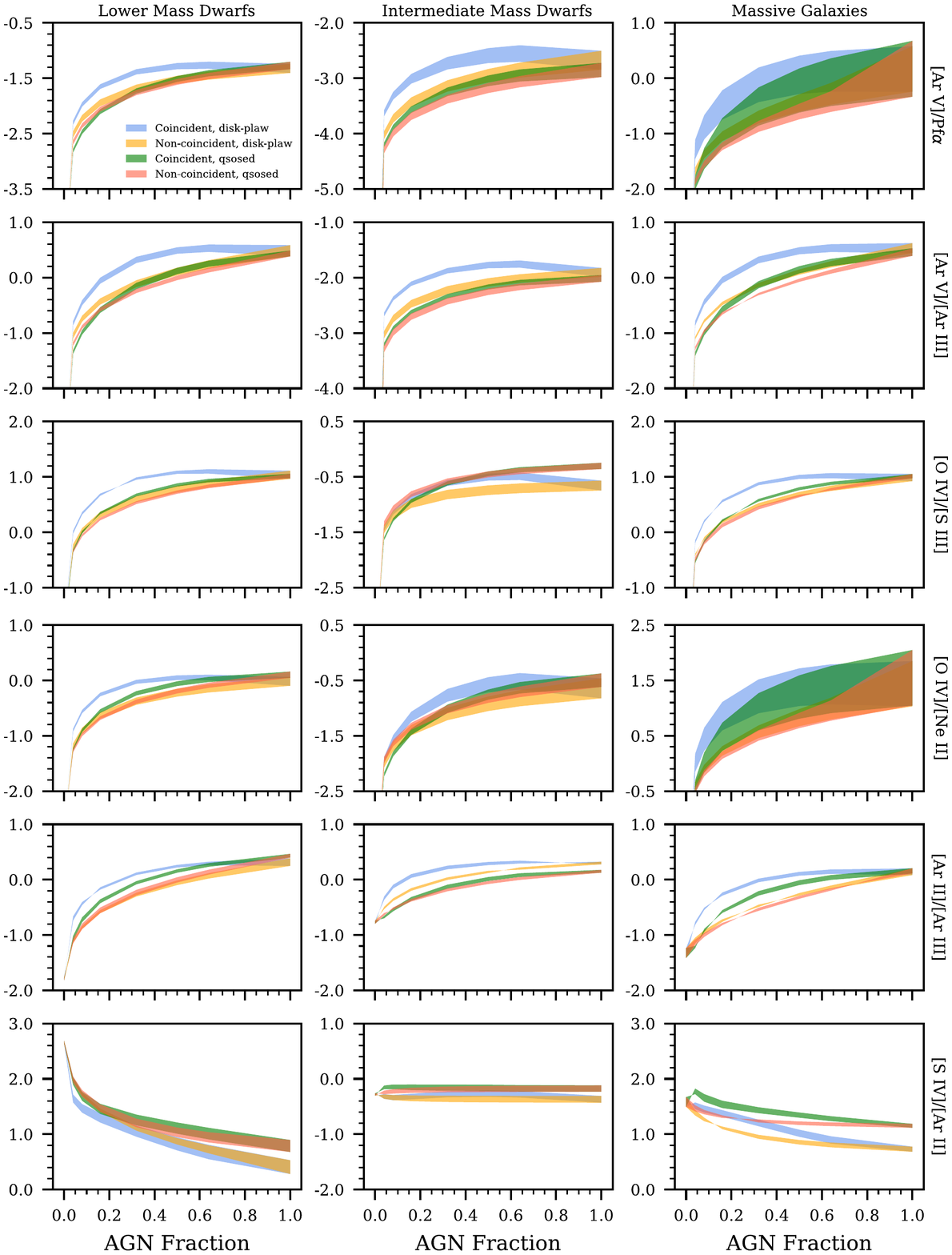}
\caption{Additional mid-IR emission line excitation diagnostics in the same format as Figure \ref{fig:AGN-frac-IR_M-5_0} with $M_{\mathrm{BH}} = 10^5~M_{\odot}$.
\label{fig:appendix_IR_M-5_0}}
\end{figure*}

In this section, we present the line ratio sensitivity for additional AGN diagnostics. Figures \ref{fig:appendix_optical_M-3_0} and \ref{fig:appendix_optical_M-5_0} display diagnostics in the optical for $M_{\mathrm{BH}} = 10^3~M_{\odot}$ and $M_{\mathrm{BH}} = 10^5~M_{\odot}$, respectively, and similarly in the mid-IR for Figures \ref{fig:appendix_IR_M-3_0} and \ref{fig:appendix_IR_M-5_0}. The columns of each figure represent a given galaxy mass, while the rows represent a given line ratio. As with the diagnostics presented in the main body of the paper, the uncertainty introduced from the AGN SED shape, cloud geometry, and mixing methodology, decreases at higher black hole mass.

In the optical, most of the emission line ratios poorly trace $f_{\mathrm{AGN}}$. It is noteworthy that [N~II]/H$\alpha$ remains relatively insensitive to AGN activity in most conditions, which combined with the line ratio's metallicity sensitivity makes it an overall poor diagnostic for dwarf AGN. In contrast, He~II/H$\beta$ and [O~I]/[O~III] are promising diagnostics for dwarf AGN. However, the wind from WR stars can contaminate He~II (\citealt{Brinchmann2008}), therefore line ratios involving [O~I] are more reliable.

In the mid-IR, one also needs to be judicious in selecting AGN sensitive diagnostics. For a $10^3~M_{\odot}$ black hole, [Ar~V]/Pf$\alpha$, [Ar~V]/[Ar~III], and [O~IV]/[Ne~II] suffer from large amounts of uncertainty for most galaxy masses regimes. In contrast, [O~IV]/[S~III] and [Ar~II]/[Ar~III] provide suitable backups for [O~IV]/Pf$\alpha$ and [Ar~II]/Pf$\alpha$ (Figures \ref{fig:AGN-frac-IR} and \ref{fig:AGN-frac-IR_M-5_0}) across the black hole masses and galaxy masses we have considered.

\end{document}